\documentclass[preprint,trackchanges]{aastex63}

\usepackage{CJK}
\usepackage{amsmath,amstext, amssymb}
\usepackage[T1]{fontenc}
\usepackage{apjfonts} 
\usepackage{graphicx}
\usepackage{color}
\usepackage{tablefootnote}
\usepackage{subfigure}
\usepackage{natbib}
\usepackage{bm}

\shorttitle{Measuring black hole masses from TDEs}
\shortauthors{Zhou et al.}

\begin{document}
\begin{CJK*}{UTF8}{gbsn}

\title[]{Measuring black hole masses from tidal disruption events and testing the $M_{\rm BH}$--$\sigma_*$ relation}

\correspondingauthor{F.K. Liu}
\email{fkliu@pku.edu.cn}

\author[0000-0002-1427-4964]{Z.Q. Zhou}
\affiliation{Department of Astronomy, School of Physics, Peking University, Beijing 100871, China}

\author{F.K. Liu$^*$}
\affiliation{Department of Astronomy, School of Physics, Peking University, Beijing 100871, China}
\affiliation{Kavli Institute for Astronomy and Astrophysics, Peking University, Beijing 100871, China}

\author{S. Komossa}
\affiliation{Max Planck Institut f\"ur Radioastronomie, Auf dem H\"ugel 69, 53121 Bonn, Germany}

\author{R. Cao}
\affiliation{Department of Astronomy, School of Physics, Peking University, Beijing 100871, China}

\author{L.C. Ho}
\affiliation{Kavli Institute for Astronomy and Astrophysics, Peking University, Beijing 100871, China}
\affiliation{Department of Astronomy, School of Physics, Peking University, Beijing 100871, China}

\author{Xian Chen}
\affiliation{Department of Astronomy, School of Physics, Peking University, Beijing 100871, China}
\affiliation{Kavli Institute for Astronomy and Astrophysics, Peking University, Beijing 100871, China}

\author{Shuo Li}
\affiliation{National Astronomical Observatories, Chinese Academy of Sciences, Beijing 100012, China}
\affiliation{Department of Astronomy, School of Physics, Peking University, Beijing 100871, China}

\begin{abstract}
   Liu and collaborators recently proposed an elliptical accretion disk model for tidal disruption events (TDEs). 
   They showed that the accretion disks of optical/UV TDEs are large and highly eccentric and suggested 
   that the broad optical emission lines with complex and diverse profiles originate in a cool eccentric 
   accretion disk of random inclination and orientation. In this paper, we calculate the radiation efficiency 
   of the elliptical accretion disk and investigate the implications for observations of TDEs. We compile  
   observational data for the peak bolometric luminosity and total radiation energy after peak brightness 
   of 18 TDE sources and compare these data to the predictions from the elliptical accretion disk model. 
   Our results show that the observations are consistent with the theoretical predictions and that the majority of the 
   orbital energy of the stellar debris is advected into the black hole (BH) without being converted into radiation.
   Furthermore, we derive the masses of the disrupted stars and the masses of the
   BHs of the TDEs. The BH masses obtained in this paper are also consistent with those calculated 
   with the $M_{\rm BH}$--$\sigma_*$ relation. Our results provide an effective method for measuring the masses 
   of BHs in large numbers of TDEs to be discovered in ongoing and next-generation sky surveys, 
   regardless of whether the BHs are located at the centers of galactic nuclei or wander in disks and halos.
\end{abstract}

\section{Introduction} \label{sec:intro}

A star would be tidally disrupted \citep{Hills1975,Rees1988,Evans1989} when it is scattered close to the vicinity of a 
supermassive black hole \citep[BH;][]{Magorrian1999,Wang2004,Chen2008,Chen2009,Liu2013,Li2017}. After the tidal 
disruption, about half of the stellar debris loses orbital energy and becomes bound to the BH. The bound stellar debris 
returns to the orbital pericenter of the progenitor star and forms an accretion disk around the BH. In the canonical 
model for such a tidal disruption event (TDE), the returned material streams are assumed to be circularized rapidly 
due to strong relativistic apsidal precession, and as a result, the accretion disk has a size of about twice the orbital
pericenter of the star \citep{Rees1988}. In this scenario, the accretion disk of a TDE is an outer-truncated analog of 
an accretion disk of an active galactic nucleus (AGN) or Galactic X-ray binary, and it radiates mainly in the soft X-rays.
The radiation in the optical and UV wave bands is expected to be rather weak and to 
decay with time as a power law much shallower than the fallback rate of the stellar debris \citep{Strubbe2009}. 
No strong broad optical emission lines are expected from such a hot accretion disk \citep{Bogdanovic2004}. 

The non-jetted TDEs discovered in the X-rays are broadly consistent with the above predictions, 
but those discovered in the optical/UV wave bands challenge the canonical model \citep[][for a recent 
review]{Komossa1999,Gezari2006,vanVelzen2011,Liu2014,Komossa2015}. Most TDEs and candidate TDEs 
discovered in the optical/UV wave bands emit radiation mainly in optical/UV wave bands and little or no 
radiation in soft X-rays \citep[e.g.,][]{Komossa2008b,Gezari2012,Wang2012,Holoien2014,Holoien2016,
Blagorodnova2019,Leloudas2019,vanVelzen2020}. Their optical/UV luminosities unexpectedly follow the fallback rate of the 
stellar debris \citep{Gezari2012,Arcavi2014,Hung2017,Wevers2017,Mockler2019}. The radiated 
energy and the implied accreted material onto the BH or the implied mass of the star could be orders of magnitude lower than expected for the tidal disruptions of main-sequence stars or 
brown dwarfs \citep{Li2002,Halpern2004,Komossa2004,Esquej2008,Gezari2008,Gezari2009,Cappelluti2009,
Maksym2010,Gezari2012,Chornock2014,Donato2014,Holoien2014,Liu2014,Holoien2016,Holoien2016a,
Blagorodnova2017,Hung2017,Saxton2017,Saxton2018,Mockler2019}. Most optical/UV TDEs and candidate TDEs
have strong broad optical emission lines with complex, asymmetric, and diverse profiles 
\citep{Komossa2008b,Gezari2012,Wang2012,Arcavi2014,Holoien2014,Holoien2016,Holoien2016a,Holoien2019a}.

It has recently been suggested that strong winds form during the phase of super-Eddington accretion, and that 
the soft X-ray radiation emitted by the accretion disk is absorbed and reprocessed into the UV band 
by the optically thick wind envelopes \citep[e.g.,][]{Dai2018}.  The broad optical emission lines are powered 
by the soft X-rays and form in the surface layers of the optically thick envelopes \citep{Roth2016}. However, 
strong outflows may cause the observed light curve to diverge significantly from the fallback rate of stellar
debris \citep{Strubbe2009,Lodato2011,Metzger2016}. 

Analytic and hydrodynamic simulations of stellar tidal disruptions 
\citep{Rees1988,Evans1989,Kochanek1994,Rosswog2009,Hayasaki2013,Hayasaki2016a,Guillochon2014,
Dai2015,Piran2015,Shiokawa2015,Bonnerot2016,Sadowski2016,Bonnerot2019a} indicate that the bound stellar 
streams circularize mainly due to the self-interaction of the streams returning at different times
caused by general relativistic apsidal precession.  Rapid formation of the accretion disk happens only
in TDEs with orbital pericenter $r_{\rm p} \lesssim 10 r_{\rm g}$, with $r_{\rm g}$ the gravitational radius
\citep{Dai2015,Shiokawa2015,Bonnerot2016,Hayasaki2016a}. However, the tidal disruption radius of a solar-type 
star by a BH of mass $\la 10^{6.5} \,M_\odot$ is $\ga 10 r_{\rm g}$, so that the general relativistic apsidal 
precession of bound streams of most TDEs is expected to be inefficient in circularizing the streams 
\citep{Shiokawa2015}. Inspired by the hydrodynamic simulations of TDEs by \citet{Shiokawa2015}, 
\citet{Piran2015} proposed that the observed luminosities of the optical/UV TDEs are powered by the orbital kinetic 
energy that is liberated by the self-crossing shocks at apocenter during the formation of the accretion disk rather than the
energy released during the subsequent accretion of matter onto the BH. Because the self-collision of  streams due to 
general relativistic apsidal precession occurs at nearly the apocenter of the most-bound stellar debris, the emission 
region could be much larger than that in the canonical circular disk model. The observed luminosities and 
temperatures near peak brightness are roughly consistent with the expectations of the self-crossing shock model 
\citep{Piran2015,Mockler2019}, provided that the orbit of the fallback material is parabolic, with the specific 
bound energy much lower than that of the most-bound stellar debris, and so long as the dissipated kinetic energy 
is efficiently converted into radiation. As noted in the original work of \citet{Piran2015}, a challenging 
question of the collision-shock model is where the energy goes that is liberated during the subsequent 
accretion of matter onto the BH. It is argued that the emissions during the formation of the accretion disk may 
dominate the radiation of TDEs because the eccentricity of the accretion disk may remain or even increase due 
to the efficient outward transfer of angular momentum by the self-crossing shocks and/or the magnetic stresses 
at apocenter, so that the gas pericenter could be reduced to the marginally stable orbit of the BH with little 
decrease in semimajor axis \citep{Svirski2017,Chan2018}. 

It is generally believed that double-peaked broad Balmer emission lines in AGNs originate from their accretion disks 
\citep{Chen1989,Chen1989a,Storchi-Bergmann1993,Storchi-Bergmann2017,Eracleous1994,Eracleous2003,Ho2000,
Shields2000,Strateva2003,Popovic2004}. By modeling the double-peaked H$\alpha$ emission line of the optical/UV 
TDE PTF09djl, \citet{Liu2017} showed that the accretion disk is extended and extremely eccentric. The 
extreme eccentricity is determined jointly by the elliptical orbit of the most-bound stellar debris and the  
self-intersection of streams. Liu and collaborators further showed that the elliptical accretion disk model can also 
explain the broad optical emission lines in the TDE ASASSN-14li, and that the diversity and time variation of its lines are 
caused by the different inclination and orientation of the elliptical disk that is precessing due to the Lense--Thirring effect 
\citep{Cao2018}. Because of their large semimajor axis and extreme eccentricity, elliptical accretion disks 
have low conversion factors of matter into radiation \citep{Liu2017, Cao2018}, consistent with the radiation 
efficiencies obtained from the analysis of the light curves of PTF09djl and ASASSN-14li \citep{Mockler2019}. The 
expected peak energy luminosities from the elliptical disk model are also well consistent with the observations of 
PTF09djl and ASASSN-14li \citep{Liu2017,Cao2018}. 

Here we further investigate the radiation efficiency of an elliptical accretion disk with a nearly uniform 
orbital eccentricity of the fluid elements in the disk plane, and we examine the implications for observations of TDEs.
We assume that the outflows from the collision shocks during the formation of the accretion disk and from the 
surface of the elliptical accretion disk, if any, are a small fraction of the fallback stellar debris. Because we are 
interested in the total radiation efficiency, we do not distinguish between the emission of radiation from the collision 
shocks during the formation of the disk and from the subsequent accretion onto the disk.  
Within the framework of the elliptical accretion 
disk model, we calculate the conversion efficiency of matter into radiation of TDEs during the accretion of matter 
onto BHs, which in the literature is always assumed to be a free parameter in modeling the luminosities
of TDEs \citep[e.g.,][]{Liu2014,Mockler2019}. With the radiation efficiency, we can calculate the expected 
peak luminosity and total radiation energy of TDEs and compare the expectations with the observations of
non-jetted TDEs in the literature. We show that the peak luminosity and the total radiation energy
expected from an elliptical accretion disk are well consistent with the observations of the non-jetted TDEs and 
candidate TDEs. 

In the elliptical accretion disk model for a Schwarzschild BH, the radiation efficiency is not a constant but depends 
significantly on the mass of the BH and the mass of the star. With the observations of the peak luminosity and the 
total radiation energy of non-jetted TDEs, we can derive the masses of the BHs and the disrupted stars. 
This provides a potentially promising method for constraining the masses of tidally disrupted stars in distant galaxies. 
Because the masses of BHs in galactic nuclei can be estimated from well-known correlations between BH mass and
host galaxy properties, we show that the BH masses obtained in this paper are consistent with those calculated 
from the $M_{\rm BH}$--$\sigma_*$ relation. This paper provides an effective technique to weigh both the 
BHs and the  stars disrupted by them, regardless of whether they hide deep in the  center of galactic nuclei and globular
clusters or wander around the galactic disk. 

The paper is organized as follows. Section~\ref{sec:disk} presents the elliptical disk model for TDEs and calculates the 
radiation efficiency. Section~\ref{sec:LEdigram} gives the peak  luminosity and the total radiation energy after peak. 
In Section~\ref{sec:obs} we compile the observational data of the peak bolometric luminosity and the total radiation
energy after peak brightness of 18 non-jetted TDEs and compare the observations with the predictions of 
the elliptical disk model. In Section~\ref{sec:bhmass} we calculate the masses of BHs and stars according to the peak 
luminosity and the total radiation energy after peak, and compare the results with those obtained from the correlation 
between BH mass and bulge properties. Discussion is presented in Section~\ref{sec:dis}, and conclusions can be found in 
Section~\ref{sec:con}.

\section{Elliptical accretion disk and radiation efficiency of TDEs}
\label{sec:disk}

We begin our calculation of the radiative efficiency by introducing our elliptical accretion disk model of
TDEs. A star is tidally destroyed when it passes by a supermassive BH with an orbital pericenter $r_{\rm p}$ smaller 
than the tidal disruption radius,
\begin{equation}
r_{\rm t} = R_* \left(M_{\rm BH} / M_*\right)^{1/3} \simeq 23.557 r_* m_*^{-1/3} M_6^{-2/3} r_{\rm S},
\label{eq:rt}
\end{equation}
where $M_{\rm BH} = 10^6  M_{\rm 6} \,M_\odot$ is the mass of the BH, $r_{\rm S}$ is the BH Schwarzschild radius, 
and $R_*= r_* \,R_\odot$ and $M_*=m_* \,M_\odot$ are the stellar radius and mass, respectively. 
Hydrodynamic simulations of TDEs show that the tidal disruption radius depends on both 
the internal structure of the star and general relativistic effects of the BH \citep{Guillochon2013,Ryu2019, Ryu2019a}. 
The result given by Equation~(\ref{eq:rt}) is an approximation with an uncertainty of order unity. For a star with an 
orbital penetration factor $\beta = r_{\rm t}/r_{\rm p}$ and a BH of mass $M_{\rm BH} = 10^6  M_{\rm 6} \, M_\odot$, 
the physical tidal disruption corresponds to $\beta\sim 0.9$ for low-mass main-sequence stars and $\beta\sim 2$ 
for stars with $M_* > 1 \, M_\odot$ \citep{Guillochon2013,Ryu2019a}. In this paper, if needed, we calculate the 
radius with the mass-radius relation $R_* \simeq R_\odot (M_*/M_\odot)^{1-\zeta}$ for main-sequence stars, where 
$\zeta \simeq 0.21$ for stellar masses $0.08 \,M_\odot < M_* \leqslant 1 \,M_\odot$ and  $\zeta \simeq 0.44$ for 
$1 \,M_\odot < M_* < 150 \,M_\odot$ \citep{Kippenhahn2012}. In the literature, hydrodynamic simulations of tidal 
disruptions have mainly been made with main-sequence stars. Here we extrapolate the results of 
main-sequence stars to brown dwarfs (BDs) using the same polytropic index. We note that the results with BDs have 
larger uncertainties.  For BDs with mass $0.01 \,M_\odot < M_* \leqslant 0.07 \,M_\odot$, we use the mass-radius 
relation appropriate for an age $t=5\, {\rm Gyr}$, $R_* \simeq 0.06 \,R_\odot (M_*/M_\odot)^{1-\zeta}$ with $\zeta 
= 9/8 $ \citep{Chabrier2000}. For BDs with mass $0.07 \,M_\odot < M_* < 0.08 \,M_\odot$, we adopt a bridge relation 
$R_* = 0.136 \, R_\odot (M_*/0.08\,M_\odot)^{1-\zeta}$ with  $\zeta = -2.637$. 

After tidal disruption,  the bound stellar debris returns to the pericenter of the progenitor star and forms an accretion 
disk mainly due to the shock produced by the collision between the post-pericenter outflowing and the freshly inflowing 
streams that results from relativistic apsidal precession \citep{Evans1989,Kochanek1994,Hayasaki2013,Dai2015,Shiokawa2015,Bonnerot2016,Hayasaki2016a}. 
The location of the collision and the conservation of angular momentum together determine the semimajor axis of the disk
\citep{Dai2015,Liu2017,Cao2018},
\begin{eqnarray}
a_{\rm d}  &\simeq & {2  r_{\rm p}  \over 2\delta  +  \sin^2(\Omega/2)} 
\simeq {2  \beta^{-1} r_{\rm t}  \over 2\delta   +  \sin^2(\Omega/2)}, 
\label{eq:dsize}
\end{eqnarray} 
as well as the eccentricity,  
\begin{eqnarray}
e_{\rm d} &\simeq & \left[1-  {(1-e_{\rm mb}^2) a_{\rm mb}\over a_{\rm d}}\right]^{1/2} \simeq   
      \left[1 - 2 \delta - \sin^2\left({\Omega \over 2}\right)\right]^{1/2}  \simeq \left[1 - 2 \delta (1+ \Delta)
      \right]^{1/2},
\label{eq:decc}
\end{eqnarray} 
with $\Delta= \sin^2 ({\Omega /2})/(2\delta)$. In Equation~(\ref{eq:decc}), $a_{\rm mb} \simeq r_{\rm t}^2 /2 
R_* $ and $e_{\rm mb} = 1- \delta$ with $\delta \simeq 2 R_* r_{\rm p} /r_{\rm t}^2 \simeq 0.02 \beta^{-1} 
m_*^{1/3} M_6^{-1/3}$ are the orbital semimajor axis and eccentricity, respectively, of the most-bound stellar 
debris. The instantaneous de~Sitter precession at periapse of the most-bound stellar debris is
\begin{eqnarray}
\Omega \simeq {6 \pi G M_{\rm BH} \over  c^2 (1- e_{\rm mb}^2) a_{\rm mb} } \simeq 
        {3\pi r_{\rm S} \over  (1+ e_{\rm mb}) r_{\rm p} } \simeq  {3\pi \over 2-\delta} \beta {r_{\rm 
        S} \over r_{\rm t}} .
\end{eqnarray}  
For a star with an orbital pericenter of $r_{\rm p} < 10 r_{\rm g}$, the 
relativistic apsidal precession of the bound stellar debris becomes important and would significantly reduce the 
eccentricity of the accretion disk.

Modeling of the double-peaked and/or asymmetric broad optical emission lines of TDEs implies that the accretion 
disks of TDEs are highly eccentric and that the eccentricity remains nearly unchanged over the disk 
\citep{Liu2017,Cao2018}. This result suggests that stellar streams circularize inefficiently. Analytical arguments as well 
as numerical hydrodynamic simulations  also show that the accretion disks of TDEs can be highly eccentric 
\citep{Guillochon2014,Shiokawa2015,Barker2016,Sadowski2016,Svirski2017,Chan2018,Ogilvie2019,
Andalman2020,Bonnerot2019a}. Recent global general relativistic hydrodynamic simulations indicate that the average 
eccentricity of TDE disks can be $ e\simeq 0.88$ at late times \citep{Andalman2020}, 
consistent with the modeling of the observed line profiles \citep{Liu2017,Cao2018}. 
Following \citet{Liu2017} and \citet{Cao2018},
and for simplicity, we assume that the eccentricity of the elliptical accretion disks of TDEs is nearly uniform over the 
disk, namely $e(a) \simeq e_{\rm d}$.  To describe the motion of the particles of highly eccentric orbits in the field of 
a Schwarzschild BH, we adopt the generalized Newtonian potential in the low-energy limit \citep[gNR;][]{Tejeda2013},
\begin{equation}
\Phi_{\rm G} = -{G M_{\rm BH} \over r} - \left({2r_{\rm g} \over r - 2 r_{\rm 
g}}\right) \left[\left({r-r_{\rm g} \over r- 2 r_{\rm g}}\right) v_{\rm r}^2 + {1\over 2} v_{\phi}^2 \right] ,
\label{eq:gNP}
\end{equation}
where $r_{\rm g}$ is the gravitational radius, $v_{\rm r}$ is the radial velocity, and $v_\phi$ is the
azimuthal velocity. In gNR, the specific binding energy $e_{\rm G}$ and 
angular momentum $l_{\rm G}$ of an elliptical orbit with semimajor axis $a$ and eccentricity $e_{\rm d}$ 
are 
\begin{eqnarray}
\epsilon_{\rm G} & \simeq & {c^2 \over 2} {r_{\rm S} [a(1-e_{\rm d}^2) - 2r_{\rm S}] \over a \left[2 a(1
      -e_{\rm d}^2) - (3+e_{\rm d}^2) r_{\rm S}\right]} 
\label{eq:eng}
\end{eqnarray}
and
\begin{eqnarray}
l_{\rm G} & \simeq & {(1-e_{\rm d}^2) a c\sqrt{r_{\rm S}} \over \sqrt{2 (1-e_{\rm d}^2) a - (3 + e_{\rm 
      d}^2)r_{\rm S}}},
\label{eq:ang} 
\end{eqnarray}
respectively \citep{Liu2017,Cao2018}. 

When the disk fluid elements migrate inward until the pericenter reaches the marginally stable radius $r_{\rm ms}$, the 
matter passes through $r_{\rm ms}$ and falls freely onto the BH. We adopt the innermost elliptical orbit of the fluid 
elements as the inner edge of the elliptical accretion disk and take the zero-torque inner boundary condition.
It has been argued that if the magnetohydrodynamic turbulence around the pericenter develops in the usual way, 
the viscous time of the elliptical accretion disk would be very long and the accretion of matter onto the BH would 
be delayed \citep{Shiokawa2015}. However, the magnetorotational instability may develop differently in an 
eccentric accretion disk \citep{Chan2018}. Both the shocks due to relativistic apsidal precession and the magnetic 
stresses near apocenter can transport angular momentum outward and efficiently reduce the gas pericenter 
\citep{Bonnerot2017b,Svirski2017}. \citet{Nealon2018} propose that gas accretion at early times
can be produced by angular momentum associated with the \citet{Papaloizou1984} instability.
If the accretion time of an elliptical accretion disk is short compared to the evolutionary time of the TDE, 
the radiative efficiency can be calculated from the energy liberated by the particles on the elliptical orbit of the 
inner edge. 

An elliptical accretion disk of constant eccentricity $e_{\rm d}$ has an inner edge $a_{\rm in} = r_{\rm ms}/(1-
e_{\rm d})$. For particles on circular orbits, the marginally stable circular orbit or the innermost stable circular orbit 
(ISCO) is at $r_{\rm ms} =3 r_{\rm S}$, while for particles with parabolic orbits, the marginally stable radius is at 
$r_{\rm ms} = 2 r_{\rm S}$. Particles with trajectories of eccentricity $0< e < 1$ are characterized by $2 r_{\rm S} < 
r_{\rm ms} < 3 r_{\rm S}$. Provided $r_{\rm ms}$, Equation~(\ref{eq:eng}) gives the specific binding energy 
$\epsilon_{\rm in}$ of particles at the inner edge of the elliptical disk, and the conversion efficiency of matter 
into radiation is
\begin{equation}
\eta = {\Delta \epsilon_{\rm G} \over c^2} =  {\epsilon_{\rm in} - \epsilon_{\rm init} \over c^2} 
        \simeq {\epsilon_{\rm in} \over 
        c^2} \simeq {1 \over 2} {r_{\rm S} [(1+e_{\rm d}) r_{\rm ms} - 
          2 r_{\rm S}] (1-e_{\rm d}^2) \over (1+ e_{\rm d}) r_{\rm ms} [2 (1+e_{\rm d}) r_{\rm ms} - (3+e_{\rm 
          d}^2) r_{\rm S}]}  ,
\label{eq:teta}
\end{equation}
where $\epsilon_{\rm init}$ is the initial specific binding energy of the inflowing bound stellar debris. Because the 
initial specific binding energy of the inflowing stellar debris is $0< \epsilon_{\rm init} \leqslant \epsilon_{\rm mb} \ll 
\epsilon_{\rm in}$, with $\epsilon_{\rm mb} \simeq {G M_{\rm BH} R_*/ r_{\rm t}^2}$ the specific binding energy 
of the most-bound stellar debris, we have $\Delta \epsilon_{\rm  G} = \epsilon_{\rm in} - \epsilon_{\rm init} 
\simeq \epsilon_{\rm in}$. For $e_{\rm d}=1$ and $r_{\rm ms} = 2r_{\rm S}$, Equation~(\ref{eq:teta}) gives $\eta
= 0$, as expected for a parabolic orbit. For $e_{\rm d} = 0$ and $r_{\rm ms} = 3r_{\rm S}$, Equation~(\ref{eq:teta}) 
gives $\eta\simeq 1/18$,  which is about $2.9\%$ smaller than the exact value of $\eta \simeq 0.0572$. From 
Equations~(\ref{eq:teta}) and (\ref{eq:decc}), we have 
\begin{eqnarray}
\eta & \simeq & \eta_0 (1-e_{\rm d}^2) \simeq \eta_0 \left[2\delta + \sin^2\left({\Omega\over 
   2}\right)\right] =2 \eta_0 \delta (1+ \Delta),
\label{eq:teff}
\end{eqnarray}
with $\Delta = \sin^2\left({\Omega/2}\right) / 2\delta$, where 
\begin{equation}
\eta_0 = {1 \over 2} {r_{\rm S} [(1+e_{\rm d}) r_{\rm ms} - 2 r_{\rm S}] \over (1+ e_{\rm d}) r_{\rm ms} 
        [2 (1+e_{\rm d}) r_{\rm ms} - (3+e_{\rm d}^2) r_{\rm S}]} 
\end{equation}
depends very weakly on both $r_{\rm ms}$ and $e_{\rm d}$.  For $e_{\rm d} = 1$ and $r_{\rm ms}=2r_{\rm S}$,
$\eta_0 = 0.0625$, while $\eta_0 = 0.0556$ for $e_{\rm d} = 0$ and $r_{\rm ms}= 3r_{\rm S}$. Adopting an average of $\eta_0 =0.059$ yields a good approximation for calculating the  radiation
efficiency with Equation~(\ref{eq:teff}). In particular, for $r_{\rm p}\gg r_{\rm S}$ we have 
\begin{equation}
\eta\simeq2.36\times10^{-3}\left({\eta_0\over0.059}\right)\left(1+\Delta\right)\beta^{-1} m_*^{1/3} M_6^{-1/3} 
\end{equation}
and $\Delta \simeq 0.25\beta^{3} r_*^{-2} m_*^{1/3} M_6^{5/3}$.

To estimate $\eta_0$ with higher accuracy, we calculate the marginally stable radius
$r_{\rm ms}$ with the eccentricity given by Equation~(\ref{eq:decc}). From Equation~(\ref{eq:ang}), 
the angular momentum of the fluid elements of the disk inner edge with eccentricity $e_{\rm d}$ is 
approximately 
\begin{equation}
l_{\rm in} \simeq {(1+e_{\rm d}) r_{\rm ms} c \sqrt{r_{\rm S}} \over \sqrt{2 (1+e_{\rm d}) r_{\rm ms} 
        - (3 + e_{\rm d}^2)r_{\rm S}}} .
\label{eq:angin}
\end{equation}
For fluid elements with specific angular momentum $l_{\rm in}$, the marginally stable radius is given by 
\begin{equation}
l_{\rm in} = l_{\rm K}
\label{eq:marpoint}
\end{equation}
\citep{Abramowicz1978}, where 
\begin{equation}
l_{\rm K} = \Omega_{\rm K} {r_{\rm ms}^3 \over r_{\rm ms}- {r_{\rm S}}}  
\label{eq:kep}
\end{equation}
is the specific angular momentum of Keplerian circular motion and $\Omega_{\rm K} = (G M_{\rm 
	BH} /r_{\rm ms}^3)^{1/2}$ is the Keplerian angular velocity. From Equations~(\ref{eq:angin})--(\ref{eq:kep}), we obtain 
\begin{equation}
r_{\rm ms} = {A \over 4 (1+e_{\rm d}) e_{\rm d}} r_{\rm S} ,
\label{eq:ms}
\end{equation}
where $A = 1+8e_{\rm d} +3 e_{\rm d}^2 + \sqrt{1+22e_{\rm d}^2 -  7 e_{\rm d}^4}$ increases 
with decreasing eccentricity. For $e_{\rm d} = 1$, Equation~(\ref{eq:ms}) give $r_{\rm ms} = 2 
r_{\rm S}$, consistent with the marginally bound radius of parabolic orbits.  When $e_{\rm d} \leqslant 0.406$, 
Equation~(\ref{eq:ms}) gives $r_{\rm ms}\geqslant 3 r_{\rm S}$. In the following calculation of the efficiency $\eta$, 
$r_{\rm ms} = 3r_{\rm S}$ is adopted for eccentricity $e_{\rm d} \leqslant 0.406$.  

\begin{figure}
\begin{center}
\includegraphics[width=0.7\columnwidth]{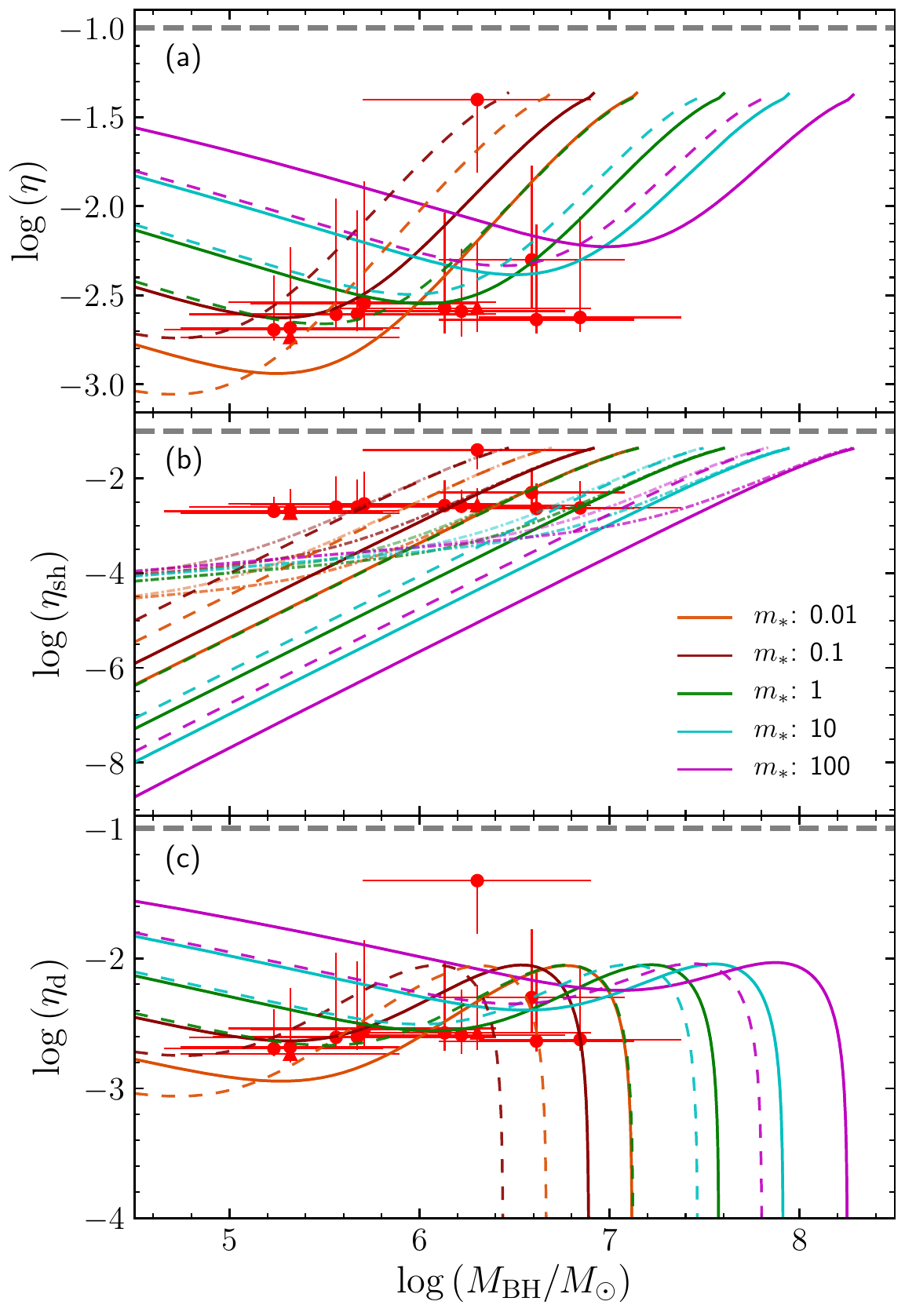}
\caption{Radiation efficiency vs. BH mass for different stellar masses for (a) the total radiation efficiency $\eta$,
(b) the radiation efficiency $\eta_{\rm sh}$ of the self-crossing shocks at apocenter, and (c) the radiation efficiency 
$\eta_{\rm d}$ during the subsequent accretion of matter onto the BH. The solid ($\beta = 1$) and dashed ($\beta 
= 2$) lines are for an initial orbital bound energy $\epsilon_{\rm init} = \epsilon_{\rm mb}$, and the dash--dotted lines 
($\beta = 1$ and $2$) in panel (b) are for an initial parabolic orbit $\epsilon_{\rm init} = 0$. Gray dashed lines are for $\eta=0.1$.
The filled circles are the radiation efficiencies and the associated uncertainties at 90\% level of the sample of TDE 
sources in Table~\ref{tab:mcmc_mbh}. The triangles are for the secondary solutions of TDEs (see 
Section~\ref{sec:starmass} for details).
\label{fig:eff}
}
\end{center}
\end{figure}

We note that in Equation~(\ref{eq:teff}), $\eta_0 \simeq 0.059$ is close to the radiation efficiency of a standard thin
accretion disk around a Schwarzschild BH, and the radiation efficiency $\eta$ of the elliptical accretion disk of TDEs 
is equivalent to the typical radiation efficiency of a standard thin accretion disk modified by a factor $(1-e^2)$. 
Equation~(\ref{eq:teff}) shows that the radiation efficiency of the elliptical accretion disk of TDEs strongly depends 
both on the masses of the BH and star and on the orbital penetration factor of the star. Figure~\ref{fig:eff} 
shows the variation in $\eta$ as a function of the masses of the BH and star for two typical orbital penetration factors, 
$\beta=1$ and $2$. Figure~\ref{fig:eff} assumes the mass-radius relations of main-sequence stars and BDs. 

The total radiation energy of the elliptical accretion disk partly comes from the self-crossing shocks near the disk 
apocenter during disk formation and partly from the subsequent accretion of matter onto the BH. Assuming that the 
heat of the shock is completely radiated away \citep[e.g.,][]{Piran2015,Wevers2019}, we can estimate the maximum
radiation efficiency $\eta_{\rm sh}$ of the self-crossing shocks as
\begin{eqnarray}
\eta_{\rm sh} &\simeq& {\epsilon_{\rm d} - \epsilon_{\rm init} \over c^2} \cr
    & \simeq & {1\over 8} { r_{\rm S} \over  r_{\rm p}} {(r_{\rm p} - r_{\rm S}) \left[2\delta +  \sin^2\left({\Omega/2}\right)\right]   \over (r_{\rm p} - r_{\rm S}) + 
    {r_{\rm S} \over 4}\left[2\delta +  \sin^2\left({\Omega/2}\right)\right]} - {\epsilon_{\rm init} \over c^2} \cr
    &\simeq & {1\over 8} { r_{\rm S} \over  r_{\rm t}} \beta
     \left[2\delta +  \sin^2\left({\Omega\over2}\right)\right] - {\epsilon_{\rm init} \over c^2} \cr
     &\simeq & 5.31\times 10^{-3} \beta r_*^{-1} m_*^{1/3}  M_6^{2/3}
     \left[2\delta +  \sin^2\left({\Omega\over2}\right)\right] - {\epsilon_{\rm init} \over c^2} \cr
     &\simeq & 2.12 \times 10^{-4} r_*^{-1} m_*^{2/3} M_6^{1/3} (1 + \Delta) - {\epsilon_{\rm init} \over c^2} .
    \label{eq:effsh}
\end{eqnarray}
In earlier works on the self-crossing shock model for optical TDEs \citep[e.g.,][]{Piran2015,Jiang2016a,Wevers2019}, an 
initial parabolic orbit is adopted for the inflowing stream so that the initial specific orbital energy is $\epsilon_{\rm init} 
= 0$. This approximation is valid for the orbits of the inflowing streams at late times, but it is inadequate for orbits 
near the peak fallback rate, whose initial orbital binding energy is similar to that of the most-bound stellar debris 
($\epsilon_{\rm init} \simeq \epsilon_{\rm mb}$). Figure~\ref{fig:eff}(b) illustrates $\eta_{\rm sh}$ for $\epsilon_{\rm init} \simeq \epsilon_{\rm mb}$ and $\epsilon_{\rm init} \simeq 0$.
The case of $\epsilon_{\rm init} \simeq 0$ (dash--dotted lines) represents the upper limits of the radiation efficiency of the self-crossing shocks at apocenter.
The radiation efficiencies of the self-crossing shocks at the peak fallback rate are expected to closely follow the curves for $\epsilon_{\rm init} \simeq \epsilon_{\rm mb}$ (solid and dashed lines).
The radiation efficiency $\eta_{\rm d}$ of the disk during the subsequent
accretion of matter onto the BH can be estimated by
\begin{eqnarray}
\eta_{\rm d}  & \simeq & \eta - \eta_{\rm sh} \cr
                            &\simeq & {\epsilon_{\rm in} - \epsilon_{\rm d} \over c^2} \cr
                          & \simeq & \left(1  - {1\over 8} { r_{\rm S} \over  r_{\rm t}} {\beta \over 
          \eta_0}\right)\eta_0  \left[2\delta +  \sin^2\left({\Omega\over 2}\right)\right] \cr
          &\simeq& \eta \left[1  - 0.09 \beta r_*^{-1} m_*^{1/3}  M_6^{2/3} \left({\eta_0 \over 0.059}\right)^{-1}
          \right]  .
\label{eq:deff}
\end{eqnarray}
Equation~(\ref{eq:deff}) gives the lower limit of the radiation efficiency during the subsequent accretion of matter
onto the BH because most of the heat energy of the self-crossing shocks at the apocenter would be converted 
back into the internal kinetic energy during the adiabatic expansion of the downstream gas \citep{Jiang2016a}.  
Equation~(\ref{eq:deff}) and Figure~\ref{fig:eff}(c) show that for BHs with $M_{\rm BH} \sim10^7 \,M_\odot$, $\eta_{\rm 
d} \sim \eta_{\rm sh}$. For BHs with $M_{\rm BH} =10^6 \,M_\odot$ and $\beta \sim 1$,  $\eta \simeq \eta_{\rm d} 
\gg \eta_{\rm sh}$, and the total luminosity is dominated by the radiation of the elliptical accretion disk.

For comparison, Figure~\ref{fig:eff} also plots the radiation efficiency $\eta =0.1$ that is typically adopted  for TDEs and 
AGNs. The total and disk radiation efficiencies change little with the initial binding 
energy of the stellar debris, while Figure~\ref{fig:eff}(b) shows that the radiation efficiency of the self-crossing shocks varies significantly. Our results are
based on the total radiation efficiency and do not change significantly with the assumption of $\epsilon_{\rm 
init}\sim 0$. The figure further shows that the total radiation efficiency $\eta$ is a convex function of BH mass,
with a minimum as low as $\eta\sim 10^{-3}$ at $M_{\rm BH} \approx 10^6 \,M_\odot - 10^7 \,M_\odot$.  The 
radiation efficiency of an elliptical accretion disk is significantly lower than $\eta = 0.1$. The radiation efficiency 
during the accretion of matter onto the BH is always significant, while the radiation efficiency $\eta_{\rm sh}$ of 
the self-crossing shocks at the time of disk formation increases with BH mass and becomes important when the 
relativistic apsidal precession is strong for high BH masses.

\section{Peak luminosities and total radiation energy after peak}
\label{sec:LEdigram}

The peak luminosities and the total radiation energies of TDEs can be 
determined observationally. We calculate these quantities according to the elliptical accretion disk model in this section and 
compare them to the observations of TDEs in the next section. 

Analytic and hydrodynamic simulations of TDEs show that the fallback rate of the 
bound stellar debris after peak can be well approximated with a power law in time 
\begin{equation}
     \dot{M} \simeq \dot{M}_{\rm p}  \left({t-t_{\rm d} \over t_{\rm p} - t_{\rm d}}\right)^{-n},
     \label{eq:accr}
\end{equation}
where $t_{\rm d}$ and $t_{\rm p}$ are the time of tidal disruption and peak mass accretion rate, respectively. In 
Equation~(\ref{eq:accr}), the power-law index of the typical value $n=5/3$ is a constant that depends on the structure
and age of the star \citep{Lodato2009,Guillochon2013,Goicovic2019,Golightly2019a,Law-Smith2019,Ryu2019a}, 
and the peak mass fallback rate 
\begin{equation}
\dot{M}_{\rm p} \simeq  A_{\gamma}  r_*^{-3/2} m_*^2  M_6^{-1/2}\, M_\odot/{\rm yr} ,
\label{eq:pmass}
\end{equation}
where $A_{\gamma}$ is a constant that depends on the penetration factor $\beta$ and the structure and age of the 
star \citep{Lodato2009,Guillochon2013,Goicovic2019,Golightly2019a,Law-Smith2019,Ryu2019a} and weakly
on the mass of the BH \citep{Ryu2019}. For full tidal 
disruptions, $n$ is typically $5/3$, especially for the fallback rate at late times \citep{Rees1988,Evans1989,Lodato2009,
Stone2013}. For partial disruptions, $n$ is well approximated with $9/4$ \citep{Guillochon2013,Coughlin2019,Ryu2019b}. 
In this work, we adopt the typical value $n=5/3$, and the results are not significantly changed with different 
values of $n$. For the tidal disruption of a solar-type star with a polytropic index $\gamma=5/3$ and a penetration 
factor $\beta=1$, we have $ n\simeq 5/3$ and $A_{5/3} \simeq 1.328$ \citep{Guillochon2013}, which we used 
to scale the peak fallback rate as required. For other stars, we use the results of hydrodynamic simulations of 
polytropic stars \citep{Guillochon2013}. We adopted $\gamma =5/3$ both for BDs with masses between $0.01 \,
M_\odot$ and $0.08 \,M_\odot$ \citep{Chabrier2000} and for low-mass main-sequence stars with masses between 
$0.08 \,M_\odot$ and $ 1 \,M_\odot$. We use $\gamma =4/3$ for high-mass stars with $M_* > 20 \,M_\odot$, 
whereas for stars with $1\,M_\odot < M_* < 20\,M_\odot$, we use a hybrid model obtained by linearly interpolating 
the results of hydrodynamic simulations of polytropes with indices $\gamma =5/3$ and $\gamma=4/3$. No 
hydrodynamic simulations of the tidal disruptions of BDs have been carried out so far, although general relativistic  
hydrodynamic simulations have been carried out for a red dwarf on an eccentric orbit \citep{Sadowski2016}.
Because the degeneracy of electron gas affects the equation-of-state of BDs and a degenerate electron gas can 
be described by polytropes of $\gamma=5/3$ \citep{Chabrier2000}, we extrapolate the results of the low-mass stars
with $\gamma =5/3$ to obtain the peak fallback rate for BDs \citep{Guillochon2013}.

For the typical radiation efficiency $\eta =0.1$ of a circular accretion disk, Equation~(\ref{eq:pmass}) 
gives the peak luminosity
\begin{eqnarray}
L_{\rm p}  &= &\eta \dot{M}_{\rm p} c^2 \cr
&\simeq& 7.53\times 10^{45} \, {\rm erg\, s^{-1}}  \, \left({\eta\over 0.1}\right) \left({A_\gamma \over 1.328}\right) M_6^{-1/2} 
      r_*^{-3/2} m_*^2 \, \cr
      &\simeq& 60 \, L_{\rm Edd} \left({\eta\over 0.1}\right)  \left({A_\gamma \over 1.328}\right) M_6^{-3/2} r_*^{-3/2} m_*^2 
\label{eq:pteff}
\end{eqnarray}
with $L_{\rm Edd} = 1.25\times 10^{44} \,M_6 \, {\rm erg\, s^{-1}}$ being the Eddington luminosity. Although 
$L_{\rm Edd}$ is independent of $\eta$, the Eddington accretion rate depends on $\eta$, as $\dot{M}_{\rm Edd} 
= L_{\rm Edd}/\eta c^2 = 0.022 \eta_{-1}^{-1}\,M_6 \,M_\odot {\rm yr^{-1}}=2.2 \eta_{-3}^{-1} \,M_6 \,M_\odot 
{\rm yr^{-1}}$, where $\eta_{-1} = \eta/0.1$ and $\eta_{-3}=\eta/10^{-3}$. For the value of $\eta = 0.1$ that is typically 
adopted for TDEs in the literature, the peak luminosity in Equation~(\ref{eq:pteff}) becomes super-Eddington for BHs 
of mass $M_{\rm BH} \la 1.5\times 10^7\,  (A_\gamma/ 1.328)^{2/3} r_*^{-1} m_*^{4/3} \,M_\odot$. Because the 
peak luminosity is highly super-Eddington, the light curves of TDEs for BHs of  $M_{\rm BH}\sim 10^6 \,M_\odot$ are 
Eddington-limited and should deviate from the fallback rate given by Equation~(\ref{eq:accr}). For a radiation 
efficiency as low as $\eta=10^{-3}$, Equation~(\ref{eq:pteff}) shows that the peak luminosity is sub-Eddington with 
$L_{\rm p}\simeq 0.6 L_{\rm Edd}$, even for $M_{\rm BH}\sim 10^6 \,M_\odot$. 

For the elliptical accretion disk with the efficiency given by 
Equation~(\ref{eq:teff}), the peak luminosity of TDEs is
\begin{eqnarray}
L_{\rm p} & = & \eta \dot{M}_{\rm p} c^2 \cr
& = & \eta_0 A_{\gamma} \left[2\delta + \sin^2\left({\Omega
	\over 2}\right)\right]  r_*^{-3/2} m_*^2 M_6^{-1/2} c^2 \,M_\odot/{\rm yr} \cr
&\simeq& 1.78\times 10^{44} \, {\rm erg\, s^{-1}}\, \left({\eta_0 \over 0.059}\right) 
\left({A_\gamma \over 1.328}\right) \beta^{-1}  r_*^{-3/2} m_*^{7/3} M_6^{-5/6} 
\times\cr
&& \left(1+0.25\beta^{3} r_*^{-2} m_*^{1/3} M_6^{5/3}\right) .
\label{eq:plum}
\end{eqnarray}
In Equation~(\ref{eq:plum}), the last equality is valid for $r_{\rm p} \gg r_{\rm S}$. Equation~(\ref{eq:plum}) 
shows that the peak luminosity of TDEs is sub-Eddington for BH masses $M_{\rm BH}\ga 10^6 \,M_\odot$.
The peak luminosity is significantly super-Eddington, and a significant Eddington-limited plateau of the peak 
brightness is expected only for those TDEs with $M_{\rm BH} \ll 1\times 10^6 \,M_\odot (\eta_0/0.059)^{6/11} 
(A_\gamma/1.328)^{6/11} \beta^{-6/11} r_*^{-9/11} m_*^{14/11}$, about an order of magnitude smaller than that 
suggested by Equation~(\ref{eq:pteff}).  When the disrupted star is a late-type M dwarf with a typical mass $m_*
\simeq 0.3$, the peak accretion rate is super-Eddington only for $M_{\rm BH} \la 5.7\times 10^5 \,M_\odot 
(\eta_0/0.059)^{6/11} (A_\gamma/1.328)^{6/11} \beta^{-6/11}$. For BHs more massive than $M_{\rm BH} 
\approx 10^6 \,M_\odot$, producing a TDE with significantly super-Eddington luminosity requires that the disrupted 
star is of B or O type. Our results predict that TDEs whose light curves clearly show Eddington-limited plateaus 
would predominantly occur in either dwarf galaxies with intermediate-mass BHs ($M_{\rm BH}\la 10^5 \,M_\odot$) 
or star-forming galaxies rich in massive O-type stars. 

\begin{figure}
\begin{center}
\includegraphics[width=0.9\columnwidth]{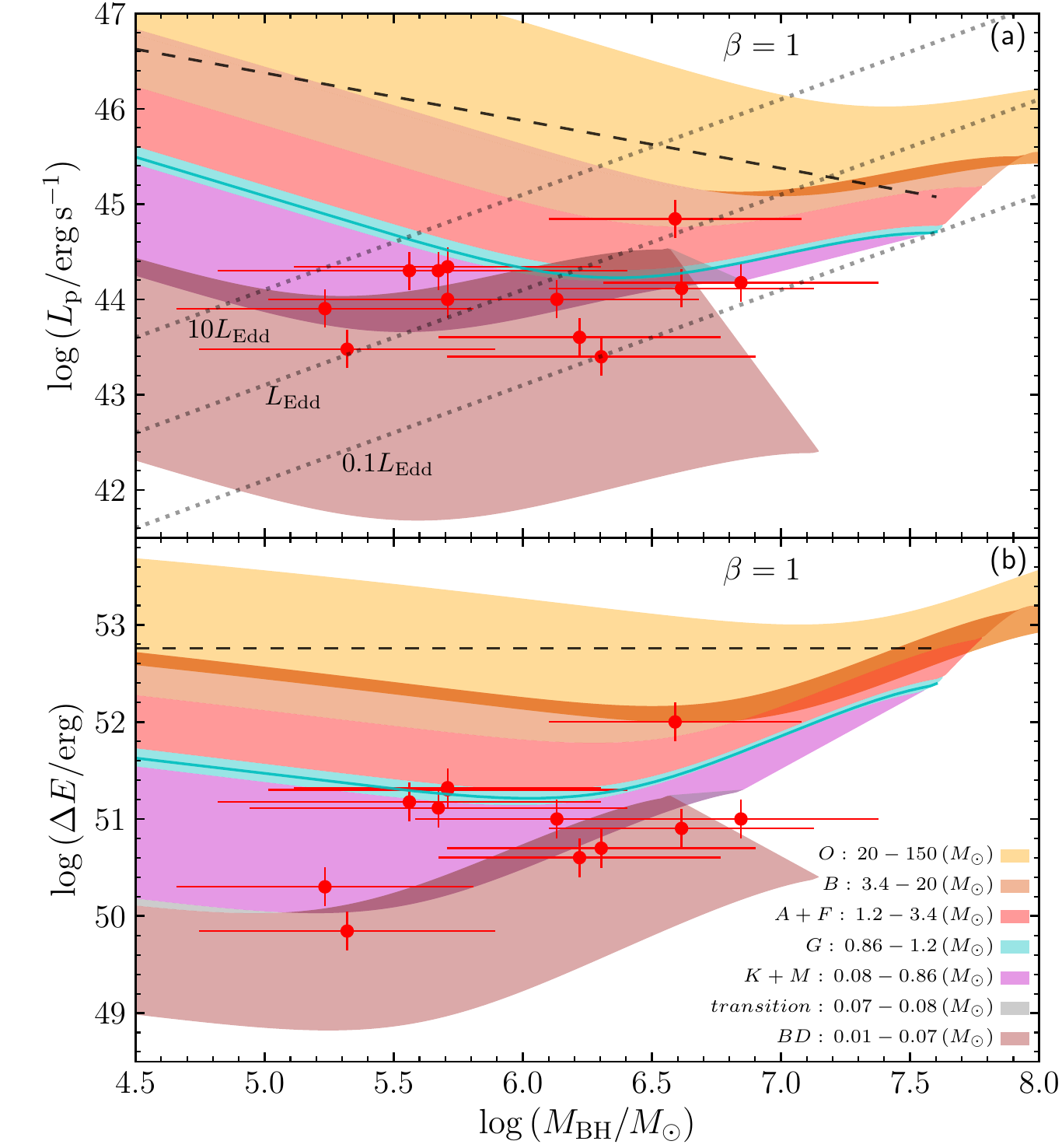}
\caption{Peak luminosity (a) and total radiation energy after peak (b) vs. BH mass.  The mass ranges of the stellar 
spectral types are adopted from \citet{Cox2000} and are color-coded. The three dotted lines in panel (a) denote 
multiples of the Eddington luminosity. The dashed lines denote the peak luminosity $L_{\rm p}$ (panel a; 
Equation~(\ref{eq:pteff})) and the total radiation energy $\Delta{E}$ (panel b) for $M_* = 1 \,M_\odot$ and 
$\eta=0.1$, while the solid lines in the cyan regions are for stellar mass $M_* = 1 \,M_\odot$ with the radiation 
efficiency $\eta$ given by Equation~(\ref{eq:teff}). The shaded areas truncate at the BH mass where $r_{\rm t} = 
r_{\rm ms}$. The filled circles show the TDE sample sources with the BH mass calculated from the $M_{\rm BH}$--$\sigma_*$ relation. 
\label{fig:LE}
}
\end{center}
\end{figure}

According to Equation~(\ref{eq:pteff}), in a circular accretion disk the peak luminosity increases with BH mass 
as $L_{\rm p} \sim L_{\rm Edd} \propto M_{\rm BH}$ for $M_{\rm BH} \la 2 \times 10^7 \,M_\odot$ because of the
Eddington limit and decreases with BH mass as $L_{\rm p} \propto M_{\rm BH}^{-1/2}$ for $M_{\rm BH} 
\ga 2\times 10^7 \,M_\odot$. A peak in the distribution of the peak luminosity is expected at $M_{\rm 
BH} \sim 2\times 10^7 \,M_\odot$. By contrast,
for an elliptical disk (Equation~(\ref{eq:plum})) $L_{\rm p} 
\propto M_{\rm BH}^{-5/6}$ for $M_{\rm BH} \la 2\times 10^6 \,M_\odot \beta^{-9/5} r_*^{6/5} m_*^{-1/5}$ and
$L_{\rm p} \propto M_{\rm BH}^{5/6}$ for $M_{\rm BH} \ga 3\times 10^6 \,M_\odot 
\beta^{-9/5} r_*^{6/5} m_*^{-1/5}$. Figure~\ref{fig:LE}a illustrates the dependence of the expected peak luminosity of an
elliptical disk as a function of BH and stellar mass for a penetration factor $\beta=1$. 
A minimum in the distribution of $L_{\rm p}$ is prominent in 
Figure~\ref{fig:LE}, and the BH mass at the minimum increases with stellar mass for main-sequence stars and decreases
with stellar mass for BDs.  We have used the mass-radius relations to obtain Figure~\ref{fig:LE}, with the mass range for different types of stars taken from \citet{Cox2000}. Figure~\ref{fig:LE} shows that the peak luminosities of TDEs strongly depend
on both the mass of the BH and the mass of the star.  They increase monotonically with star mass except at the 
transitions from BDs to late M-type main-sequence stars, and from F-type to early O-type main-sequence stars. 
Figure~\ref{fig:LE} also plots the peak luminosity as a function of BH mass given by Equation~(\ref{eq:pteff}) 
for $\eta=0.1$ and $M_*= 1 \,M_\odot$. This shows that for TDEs with
solar-type stars the peak luminosities predicted by the popular circular accretion disk model are typically orders of magnitudes higher than those of the elliptical accretion disk model.

From Equation~(\ref{eq:accr}),  the total fallback and accreted mass after peak time $t_{\rm p}$ is
\begin{eqnarray}
\Delta{M}_* & \simeq &   \int_{t_{\rm p}}^{\infty} { \dot{M}_{\rm p}  \left({t-t_{\rm d} \over t_{\rm p} 
		-t_{\rm d}}\right)^{-n} {\rm d}t}  \simeq   {\dot{M}_{\rm p} \over n-1} \Delta{t_{\rm p}}  \cr
& \simeq & {A_\gamma B_\gamma \over n-1} m_* \,M_\odot \simeq {3\over 2} A_\gamma B_\gamma
m_* \,M_\odot, 
\label{eq:accm}
\end{eqnarray}
where $\Delta{t}_{\rm p}  = t_{\rm p} -t_{\rm d} \simeq B_{\gamma} M_6^{1/2} m_*^{-1} r_*^{3/2}
\, {\rm yr}$ is the time of peak accretion and $B_\gamma$
is a constant depending on the polytropic index $\gamma$ and the orbital penetration factor $\beta$ 
of the star \citep{Guillochon2013}. To obtain the last equality in Equation~(\ref{eq:accm}), we have adopted the 
typical value $n=5/3$. For $\gamma=5/3$ and $\beta=1$, we have $B_{5/3}= 0.1618$ (see the appendix 
of \citealt{Guillochon2013}). Equation~(\ref{eq:accm}) gives the expected total accreted stellar mass 
$\Delta{M}_* \simeq 0.322(A_\gamma B_\gamma/0.215) m_*  \,M_\odot$ after the peak time of  the
light curve. For a circular accretion disk of typical $\eta =0.1$,  the total
accreted mass gives a total radiation energy of $\Delta{E} =\eta \Delta{M}_* c^2 
\simeq 5.76\times 10^{52} \, {\rm erg}\, \eta_{-1} (A_\gamma B_\gamma/0.215) m_*$, which only depends on the mass of the star in a full disruption and is independent of the BH mass. For a typical star of mass $M_*=0.3 
\,M_\odot$, the expected total radiation energy is $\Delta{E} \simeq 1.73\times 10^{52} \, {\rm erg}
\, \eta_{-1} (A_\gamma B_\gamma/0.215)$.

For the elliptical accretion disk with the radiation efficiency given by Equation~(\ref{eq:teff}), 
Equation~(\ref{eq:accm}) gives the total energy of radiation
\begin{eqnarray}
\Delta{E} &\simeq& \eta \Delta{M}_* c^2 \cr
&\simeq& {\eta_0 \over n-1} \left[2\delta + \sin^2\left({\Omega\over 2}\right)\right]
A_\gamma B_\gamma m_* M_\odot c^2 \cr
&\simeq&  1.36\times 10^{51} \, {\rm erg}\, \left({\eta_0 \over 0.059}\right) \left({A_\gamma 
	B_\gamma \over 0.215}\right) \beta^{-1}  m_*^{4/3} M_6^{-1/3}  \times \cr
&& \left(1+ 0.25 \beta^{3} r_*^{-2} m_*^{1/3} M_6^{5/3}\right)  ,
\label{eq:toteng}
\end{eqnarray}
which depends not only on the mass of the star but also on the BH mass and orbital penetration factor of the star.
For a solar-type star disrupted by a BH with a mass of $M_{\rm BH} = 10^6 \, M_\odot$ and $10^7 \,M_\odot$, the 
total radiation energy is about 34 and 7 times lower, respectively, than that expected with the typical efficiency 
$\eta=0.1$. Comparing the total energy radiated by circular versus eccentric disk models, we find that the missing 
energy problem \citep{Piran2015} may be due to the high radiative efficiency of circular disk models. This problem is absent in the elliptical accretion disk model. We further discuss this point in Section~\ref{sec:obs}. 
Figure~\ref{fig:LE}(b) presents the expected total radiation energy as a function of $M_{\rm BH}$ and $M_*$ for orbital
penetration factor $\beta=1$. We also show the expected total radiation energy computed for TDEs of solar-type
stars with $\eta=0.1$. Provided $M_*$, the expected total emitted 
energy weakly decreases with BH mass as $\Delta{E} \propto M_{\rm BH}^{-1/3}$ until $M_{\rm BH} \sim
2\times 10^6 \,M_\odot \beta^{-9/5} r_*^{6/5} m_*^{-1/5} \sim 2\times 10^6 \,M_\odot \beta^{-9/5} 
m_*^{1-1.2\zeta}$ and significantly increases when $M_{\rm BH} \ga 3\times 10^6 \,M_\odot \beta^{-9/5} 
r_*^{6/5} m_*^{-1/5} \sim 3\times 10^6 \,M_\odot \beta^{-9/5} m_*^{1-1.2\zeta}$. The transition occurs 
at $M_{\rm BH} \sim 3\times 10^6 \,M_\odot \beta^{-9/5} m_*^{1-1.2\zeta}$, with 
$\zeta=0.21$ for $0.08 < m_* \leqslant 1$ and $\zeta=0.44$ for $1 < m_* \leqslant 150$. 

From the total radiation energy given in Equation~(\ref{eq:toteng}), we can calculate the expected 
accreted stellar mass to power a TDE with the canonical radiation efficiency $\eta = 0.1$, 
\begin{eqnarray}
\Delta{M}_{\rm app} &=& {\Delta{E} \over 0.1c^2} = \left({\eta\over 0.1}\right) \Delta{M}_* \cr
& \simeq &
15 \eta_0 \left[2\delta + \sin^2\left({\Omega\over 2}\right)\right]
A_\gamma B_\gamma m_* M_\odot ,
\label{eq:accmass}
\end{eqnarray}
which for $r_{\rm p} \gg r_{\rm S}$ gives
\begin{eqnarray}
\Delta{M}_{\rm app} &\simeq& 7.61\times 10^{-3} \, M_\odot\, \left({\eta_0 \over 0.059}\right) \left({A_\gamma 
	B_\gamma \over 0.215}\right) \beta^{-1}  m_*^{4/3} M_6^{-1/3}  \times \cr
&& \left(1+ 0.25 \beta^{3} r_*^{-2} m_*^{1/3} M_6^{5/3}\right).
\label{eq:acclar}
\end{eqnarray}
Equations~(\ref{eq:accmass}) and (\ref{eq:acclar}) show that the accreted stellar mass inferred with the 
canonical $\eta=0.1$ is much lower than the actual accreted stellar mass given by
Equation~(\ref{eq:accm}). This suggests that the observed low accreted stellar mass of TDEs is due to the high radiation efficiency adopted in the literature. In the following, we call the accreted stellar mass 
inferred with $\eta = 0.1$ in Equation~(\ref{eq:accmass}) the apparent accreted stellar mass. The apparent 
accreted stellar mass after peak is $\Delta{M}_{\rm app} \simeq 0.045 \,M_\odot$ (Equation~\ref{eq:acclar}) 
or $0.04\,M_\odot$ (Equation~\ref{eq:accmass}) for $M_{\rm BH}= 10^7 \,M_\odot$ and $M_*=1\,M_\odot$. The 
apparent accreted stellar mass is $\Delta{M}_{\rm app} \simeq 3 \times 10^{-3} \,M_\odot$ and $7 \times 10^{-4} 
\,M_\odot$ for tidal disruptions of M stars with $M_* = 0.3 \,M_\odot$ and BDs with
$0.03 \,M_\odot$ by a BH of mass $10^6 \,M_\odot$, respectively.

Both the peak luminosity and the total radiation energy 
depend on BH mass, star mass, and orbital penetration factor $\beta$. The fallback rate of 
full disruption weakly depends on $\beta$ \citep{Guillochon2013,Stone2013,Ryu2019a}, suggesting that the peak 
luminosity $L_{\rm p}$ and the total radiation energy $\Delta E$ depend on $\beta$ mainly because of the radiation 
efficiency $\eta$. For a tidal disruption with $\Delta =0.25 \beta^{3} r_*^{-2} m_*^{1/3} M_6^{5/3} \la 1$, or with 
$\beta \la 2.1$ and $M_{\rm BH} \la 10^{6.5} \,M_\odot$, we have $L_{\rm p} \propto \beta^{-1} (1+\Delta) 
m_*^{(5+9\zeta)/6} M_6^{-5/6} \propto  \beta^{-1} (1+\Delta) m_*^{1.15} M_6^{-5/6}$, and the total radiation 
energy $\Delta E \propto \beta^{-1} (1+\Delta) m_*^{4/3} M_6^{-1/3}$ for $\zeta=0.21$, both of which are 
nearly independent of the orbital penetration factor because $1.2 \la \beta^{-1} (1+\Delta)  \la 1.6 $ 
(which varies by only $\sim$30\%) for $0.7 \la \beta \la 2.1$. Provided $L_{\rm p}$ and $\Delta E$, we can uniquely 
determine the masses of the star and the BH, but we can only poorly constrain the penetration factor for 
TDEs of shallow orbital pericenter 
penetration $\beta \sim 1$, which is expected for most TDEs. By comparison, for tidal disruption with $\Delta 
\gg 1$ (either $\beta\ga 2.5$,  $m_* \la 0.11$, or $M_{\rm BH} \ga 10^{6.7} \,M_\odot$), we have the peak luminosity 
$L_{\rm p} \propto \beta^2 m_*^{(-5+21 \zeta)/6} M_6^{5/6} \propto \beta^2 m_*^{-0.098} M_6^{5/6}$ and the total 
radiation energy $\Delta E \propto \beta^2 m_*^{(-1+6\zeta)/3} M_6^{4/3} \propto \beta^2 m_*^{0.087} M_6^{4/3}$, 
both strongly sensitive to the penetration factor and the BH mass, but nearly independent of the mass 
of the disrupted star. Given $L_{\rm p}$ and $\Delta E$, we can uniquely determine the mass of BH and the 
orbital penetration factor, but not the mass of the star. The above analysis shows that the BH mass of TDEs can be well determined by observing $L_{\rm p}$ and $\Delta{E}$. However, the mass of the disrupted 
star and the orbital penetration factor cannot be determined simultaneously: the mass of the disrupted star can 
be uniquely determined for TDEs with negligible relativistic apsidal precession of the most-bound stellar debris 
with $\Delta \la 1$, or the penetration factor $\beta$ can be well determined observationally if  the relativistic 
apsidal precession of the most-bound stellar debris is significant with $\Delta \gg 1$.

\section{Comparison with observations}
\label{sec:obs}

In Section~\ref{sec:LEdigram} we calculated the expected peak luminosities and total radiation energies after 
the peak of TDEs within the framework of an elliptical accretion disk model. In this section, we compare the 
model predictions to the observations of the peak luminosities and total radiation energies of TDEs. We adopt a 
${\rm\Lambda CDM}$ cosmology with $H_0=70\,{\rm km\,s^{-1}\, Mpc^{-1}}$, $\Omega_m=0.3$, and 
$\Omega_{\Lambda}=0.7$.

\subsection{The observational data}
\label{sec:sample}

\begin{deluxetable*}{ccccccccccccccc}
	\tablecaption{The Sample 
	\label{tab:obs}}
	\tablewidth{0pt}
	\tabletypesize{\scriptsize}
	\tablehead{
		\colhead{Name} & \colhead{$z$} & \colhead{Type} & \colhead{Ref.} & 
		\colhead{$\sigma_*$} & \colhead{Ref.} & \colhead{$\log(M_{\rm BH, \sigma})$} & 
		\colhead{$\log(M_{\rm tot})$} & \colhead{Ref.} & 
		\colhead{$\log(M_{\rm BH, tot})$} & \colhead{$B/T$\tablenotemark{\scriptsize{a}}}  & 
		\colhead{$L{\rm p}$} & \colhead{Ref.} & \colhead{$\Delta{E}$} & \colhead{Ref.} 
		\\
		\colhead{} & \colhead{} & \colhead{} & \colhead{} & 
		\colhead{(km s$^{-1}$)} & \colhead{} & \colhead{($M_\odot$)} &
		\colhead{($M_\odot$)} & \colhead{} & 
		\colhead{($M_\odot$)}  & \colhead{} & 
		\colhead{($10^{44}\,{\rm erg~s^{-1}}$)} & \colhead{} & \colhead{($10^{50}\, {\rm erg}$)} & \colhead{}
	} 
	\colnumbers
	\startdata
	iPTF16fnl    & 0.0163  &Opt./UV &1,2       &55$\pm$2  &3  &5.32$^{+0.57}_{-0.58}$	&9.8	&4  &$5.61 \pm 0.65$ 	&0.29       &0.3 &2	 &0.7\tablenotemark{\scriptsize{b}}	&2\\
   AT 2018dyb 	  & 0.0180  &Opt./UV & 5	    & 96$\pm$1 & 5 & 6.62$^{+0.51}_{-0.52}$ &10.08	& 5 &$6.06 \pm 0.65$ 	&0.43  &1.3	& 5	 &8\tablenotemark{\scriptsize{b}}		&5\\
   ASASSN-14li   & 0.0206  &Opt./UV &6			&78$\pm$2  &3  &6.13$^{+0.55}_{-0.55}$	&9.6	&4  &$5.29 \pm 0.65$ 	&0.22   &1	&6,7 &10\tablenotemark{\scriptsize{b}}	&6,7\\
	ASASSN-14ae   & 0.0436  &Opt./UV &8			&53$\pm$2  &3  &5.23$^{+0.58}_{-0.58}$	&9.8	&4  &$5.61 \pm 0.65$ 	&0.29  &0.8	&8	 &2\tablenotemark{\scriptsize{b}}		&8\\
	ASASSN-15oi   & 0.0479  &Opt./UV &9 		&61$\pm$7   &10 &5.56$^{+0.74}_{-0.77}$	&9.9	&4  &$5.77 \pm 0.65$ 	&0.34  &2	&11	 &15\tablenotemark{\scriptsize{b}}	&11\\
	PTF09ge      & 0.064   &Opt./UV  &12		&81$\pm$2  &3  &6.22$^{+0.55}_{-0.55}$	&10.1	&4  &$6.09 \pm 0.65$ 	&0.44  &0.4	&13	 &4\tablenotemark{\scriptsize{b}}		&13\\
	iPTF15af 	  & 0.07897 &Opt./UV &14		&106$\pm$2 &3  &6.85$^{+0.53}_{-0.53}$	&10.2	&4  &$6.25 \pm 0.65$ 	&0.47 &1.5	&14	 &10\tablenotemark{\scriptsize{b}}	&14\\
	SDSS J0952+2143 & 0.079   &Opt./UV &15		&95		   &15 &6.59$^{+0.49}_{-0.49}$	&10.37	&16 &$6.53 \pm 0.65$ 	&0.53      &7	&17	 &100\tablenotemark{\scriptsize{b}}	&17\\
	PS1-10jh   	  & 0.1696  &Opt./UV &18		&65$\pm$3  &3  &5.71$^{+0.59}_{-0.60}$	&9.5	&4  &$5.13 \pm 0.65$ 	&0.19       &2.2	&18	 &21	&18\\
	PTF09djl     & 0.184   &Opt./UV &12		   &64$\pm$7  &3  &5.67$^{+0.73}_{-0.76}$	&10.1	&4  &$6.09 \pm 0.65$ 	&0.44       &2	&13	 &13\tablenotemark{\scriptsize{b}}	&13\\
	GALEX D23H-1  & 0.1855   &Opt./UV &19	    &84$\pm$4  &10 &6.30$^{+0.60}_{-0.60}$	&10.3	&4  &$6.41 \pm 0.65$ 	&0.51      &0.25	&19	 &5\tablenotemark{\scriptsize{b}}		&19\\
	GALEX D1-9    & 0.326	&Opt./UV &20	    &65$\pm$6  &10 &5.71$^{+0.70}_{-0.72}$	&10.3	&4  &$6.41 \pm 0.65$ 	&0.51      &1	&20	 &20\tablenotemark{\scriptsize{b}}	&20\\
	\hline
	XMMSL1 J0740  & 0.0173	&X-ray   &21	    &\nodata &\nodata &\nodata &$\sim 10.62\tablenotemark{\scriptsize{c}}$	&\nodata &$6.92 \pm 0.65$ &0.60 &2 &21 &6\tablenotemark{\scriptsize{b}}	&21\\
	ASASSN-19bt   & 0.0262  &Opt./UV &22	    &{\nodata} &{\nodata} &{\nodata} &10.04	&22 &$6.00 \pm 0.65$ 	&0.41    &1.3	&22	 &10\tablenotemark{\scriptsize{b}}	&22\\
	AT 2018fyk 	  & 0.059   &Opt./UV &23	    &{\nodata} &{\nodata} &{\nodata} &10.2	&23 &$6.25 \pm 0.65$ 	&0.47    &3	&23	 &30\tablenotemark{\scriptsize{b}}	&23\\
	PS18kh 		  & 0.071	&Opt./UV &24, 25, 26  &{\nodata} &{\nodata} &{\nodata} &10.15	&24 &$6.17 \pm 0.65$ 	&0.46    &0.9	&24	 &7\tablenotemark{\scriptsize{b}}		&24\\
	AT 2017eqx 	  & 0.1089	&Opt./UV &27	    &{\nodata} &{\nodata} &{\nodata} &9.36	&27 &$4.90 \pm 0.65$ 	&0.14    &1	&27	 &4\tablenotemark{\scriptsize{b}}		&27\\
	PS1-11af 	  & 0.4046	&Opt./UV &28	    &{\nodata} &{\nodata} &{\nodata} &10.1	&4  &$6.09 \pm 0.65$ 	&0.44    &0.8	&28	 &6\tablenotemark{\scriptsize{b}}		&28\\
	\enddata
	\tablecomments{The sample sources are divided into two groups: the upper part of the table shows the sources with a measurement of the stellar velocity dispersion of the host galaxies, and the lower part shows the sources without such a measurement.}
	\tablenotetext{\scriptsize{a}}{The bulge-to-total mass ratio ($B/T$) is estimated with an empirical relation between
$B/T$ and the total stellar mass of the galaxy \citep{Stone2018}. It is obtained by averaging the $B/T$ for different total stellar
mass bins and has a very large uncertainty.}
	\tablenotetext{\scriptsize{b}}{The energy is obtained by extrapolating the observations from the period of observational campaign to infinity.}
	\tablenotetext{\scriptsize{c}}{The total stellar mass is estimated with the the 2MASS apparent $K$-band 
magnitude $K=10.96$~mag \citep{Saxton2017} and the average stellar mass-to-light ratio ($M/L$) \citep{Bell2003}. This 
is only a rough estimate, as we lack the color of the host galaxy.}
   \tablerefs{(1) \citet{Blagorodnova2017}, (2) \citet{Brown2018}, (3) \citet{Wevers2017},(4) \citet{vanVelzen2018}, (5) \citet{Leloudas2019}, (6) \citet{Holoien2016}, (7) \citet{Brown2017}, (8) \citet{Holoien2014}, (9) \citet{Holoien2016a}, (10) \citet{Wevers2019}, (11) \citet{Holoien2018}, (12) \citet{Arcavi2014}, (13) \citet{vanVelzen2019a}, (14) \citet{Blagorodnova2019}, (15) \citet{Komossa2008b}, (16) \citet{Graur2018}, (17) \citet{Palaversa2016}, (18) \citet{Gezari2012}, (19) \citet{Gezari2009}, 
   (20) \citet{Gezari2008}, (21) \citet{Saxton2017}, (22) \citet{Holoien2019}, (23) \citet{Wevers2019a}, (24) \citet{Holoien2019a}, (25) \citet{vanVelzen2019}, (26) \citet{Hung2019}, (27) \citet{Nicholl2019}, (28) \citet{Chornock2014}
	}
\end{deluxetable*}

Among the 30--60 TDEs and candidate TDEs\footnote{An {\it Open TDE Catalog} is available at 
\url{https://tde.space}} \citep{Komossa2015}, some have well-observed peaks in their light curves and can be 
used in the comparison between the model prediction and observation.  Because we are interested only in 
the energy released by the accretion disk formed from the tidal disruptions of stars, we include a TDE in the sample 
only when (1) it is not relativistically jetted, (2) the host galaxy does not show any long-term AGN activity, (3) the 
peak brightness is well detected, and (4) the location coincides with the nucleus of the host galaxy. The peak of 
the birghtness is well detected if neither the observational time gap before nor after the maximum luminosity of 
the light curve is longer than 30 days in the rest frame of the source. We adopted 30 days as the upper limit of the 
observational time gap because the time between the disruption of a solar-type star by a BH of $M_{\rm 
BH}=10^6 M_\odot$ and the peak accretion rate is $\Delta{t}_{\rm p} \simeq 59\, {\rm days}$. The fourth 
requirement ensures that the BH mass can be estimated from the empirical relation between BH mass and 
bulge stellar velocity dispersion ($M_{\rm BH}$--$\sigma_*$ relation), if a measurement of $\sigma_*$ is available. 
This latter requirement excluded from the final sample the TDE candidates ROTSE {\it Dougie} and AT 2018cow 
because they are off-nucleus, even though the peaks of the light curves have been well detected 
\citep{Vinko2015, Kuin2019, Margutti2019, Perley2019}. 

We assembled a final sample of 18 sources (Table~\ref{tab:obs}). All except ASASSN-14li and SDSS J0952+2143 have 
well-observed light-curve peaks in the wave bands of discovery. The peak brightness of TDE ASASSN-14li cannot be 
constrained in the optical/UV wave band of discovery because the observational time gap before the first detection 
of the event on 11 November 2014 is 121 days in the observer frame or 118.6~days in the source frame 
\citep{Holoien2016a}, although the peak was well detected in the soft X-rays \citep{Miller2015, Brown2017,
Bright2018}. TDE candidate SDSS~J0952+2143 was discovered through the detection of transient ultra-strong optical 
emission lines during the SDSS survey \citep{Komossa2008b} and has an unfiltered optical light curve from the Lincoln 
Near Earth Asteroid Research (LINEAR) survey \citep{Palaversa2016}. Table~\ref{tab:obs} divides the sample into two 
groups, according to whether or not stellar velocity dispersion is available for the host galaxy. 
Both luminosity-weighted and central line-of-sight velocity dispersions are measured in the literature, and there is no significant difference between them \citep{Wevers2017,Wevers2019}. The velocity dispersions are not affected significantly by the presence of disks in the host galaxies.  
Column~7 gives the BH mass obtained with the host stellar velocity dispersion in Column~5.  Extensive works on the $M_{\rm BH}$--$\sigma_*$ relation have been published in the 
literature and indicate that the $M_{\rm BH}$--$\sigma_*$ relation depends both on the type of host galaxy and on the range of the BH masses in the sample \citep[e.g.,][]{Kormendy2013}. 
Because TDEs are expected to occur in all types of galaxies, neither the $M_{\rm BH}$--$\sigma_*$ relation obtained from the early-type galaxies nor the one from the late-type galaxies could give good estimates
of the BH masses of TDEs. Therefore, we  estimate the BH masses using the $M_{\rm BH}$--$\sigma_*$ relations obtained 
from all types of galaxies \citep{vandenBosch2016,She2017}.
Because the early- and late-type galaxies have their own $M_{\rm BH}$--$\sigma_*$ relations with different 
slopes and zeropoints, the obtained $M_{\rm BH}$--$\sigma_*$ relation depends on the sample of galaxies. With the tabulated data of all types of galaxies \citep{Kormendy2013}, 
\citet{She2017} obtained the relation $\log(M_{\rm BH}/M_\odot) = (8.32\pm0.05)+(5.20\pm0.37) \log(\sigma_*/
200\,{\rm km \,s^{-1}})$ with an intrinsic scatter of 0.44~dex. Because most TDEs are expected to be caused by a BH 
of mass lower than $10^8 \,M_\odot$,  here we calculate the BH masses with the $M_{\rm BH}$--$\sigma_*$ relation, 
$\log(M_{\rm BH}/M_\odot)=(8.32\pm0.04)+(5.35\pm 0.23) \log(\sigma_*/200\,{\rm km\,s^{-1}})$ with the intrinsic 
scatter of $0.49\pm0.03$~dex \citep{vandenBosch2016}, which are consistent with the results obtained by She et al. (\citeyear{She2017}; with the difference of $\Delta\log(M_{\rm BH}/M_\odot)$ ranging from  $0.04$ for $\sigma_*=110 
\,{\rm km\,s^{-1}}$ to $0.08$ for $\sigma_*=55 \,{\rm km\,s^{-1}}$; see Figure~\ref{fig:mbh-sigma} for a detailed comparison). We note that the BH mass obtained in this way is from more low-mass objects and the sample is twice as large as the most previously studied sample. The uncertainties of the BH mass in Table~\ref{tab:obs} come from both the observational uncertainties of 
stellar velocity dispersion and the intrinsic scatter of the $M_{\rm BH}$--$\sigma_*$ relation. Although TDEs are expected to occur in all types of galaxies, the spectroscopical observations show that the host galaxies of most known TDEs 
are E+A or post-starburst galaxies \citep{Arcavi2014,French2016}. Post-starburst galaxies are in transition between 
star-forming spirals and passive early-type galaxies. No $M_{\rm BH}$--$\sigma_*$ relation specifically for 
post-starburst galaxies is available in the literature. We may estimate the BH masses of post-starburst galaxies 
by averaging the BH masses obtained separately with the $M_{\rm BH}$--$\sigma_*$ 
relations for early- and late-type galaxies. Using this method and the relations in \citet{McConnell2013}, we obtain the interpolated $M_{\rm BH}$--$\sigma_*$ relation for TDEs, 
\begin{equation}
\log(M_{\rm BH}/M_\odot) = 8.23+5.13 \log(\sigma_*/200\,{\rm km\,s^{-1}}) . 
\label{eq:mbhsigma}
\end{equation}
The last equation is nearly independent of 
the distributions of the BH masses in the subsample galaxies.

It has recently been suggested  that the BH mass may correlate with the total stellar mass of the host galaxy 
\citep{Reines2015}. The relation between the BH mass and the total galactic stellar mass obtained from the AGN 
sample has been used to estimate the BH masses of TDEs in the literature 
\citep[e.g.,][]{Gezari2017,Lin2017,Holoien2018,Leloudas2019,Saxton2019,Wevers2019}. The relation between the BH 
mass and the total  galactic stellar mass has recently been updated \citep{Greene2020}. Columns~8 and 9 list the total 
stellar masses of the host galaxies and the references, and Column~10 is the BH mass estimated with the $M_{\rm BH}$--$M_{\rm tot}$ relation, $\log(M_{\rm BH}/M_\odot)=6.70 +1.61\log(M_{\rm tot}/3\times 10^{10} \, M_\odot)$ 
\citep{Greene2020}. When more than one measurement of the total stellar mass is available in the literature, we 
adopted one of them in the calculations. For XMMSL1 J0740, no measurement of the total stellar mass is available 
in the literature. We use the 2MASS apparent $K$-band magnitude K = 10.96 \citep{Saxton2017} and adopt the 
average mass-to-light ratio ($M/L_{\rm K}$) as a function of total stellar mass from the appendix of \citet{Bell2003} 
to estimate the total stellar mass of the host of XMMSL1 J0740. Because the stellar mass-to-luminosity ratio $M/L_{\rm 
K}$ is color-based (although it is less sensitive in $K$ band) and we lack the color information, the total stellar mass 
of the host galaxy is only a rough estimate. We note that the BH mass of XMMSL1 J0740, $M_{\rm BH}\sim 
10^{6.92}\,M_\odot$, estimated from the total stellar mass, is consistent with the results of 
\citet{Saxton2017}\footnote{The measurement of the host stellar mass of XMMSL J0740 could be $3-3.5\times 
10^9 \,M_\odot$ based on private communication with Saxton. The BH mass estimated with the measurement and 
the $M_{\rm BH}$--$M_{\rm tot}$ relation is about $10^5 \,M_\odot$, about two order of magnitude lower than the 
value quoted in the table and roughly consistent with the MCMC result in Table~\ref{tab:mcmc}.}.
Because it is difficult to compute the uncertainty of the total stellar mass of the host galaxy, the uncertainty
of BH mass in Column~10 is only due to the scatter of the $M_{\rm BH}$--$M_{\rm tot}$ relation. Column~11 gives the 
ratio ($B/T$) of the bulge and total stellar mass of the host galaxy. Because the bulges of the host galaxies of most TDEs are not resolved, we estimate $B/T$ from the empirical relation between the mass ratio $B/T$ and the total stellar mass of the galaxy \citep{Stone2018}. The correlation of the mass ratio $B/T$ and the total stellar mass of the galaxy has a very large
scatter. The total stellar masses of the sample galaxies in \citet{Stone2018} are binned, and the ratio $B/T$ in Table~\ref{tab:obs} is the 
average of each bin, which, we note, has very large uncertainties.

Table~\ref{tab:obs} lists the peak bolometric luminosities.  Ideally, the peak bolometric luminosity should be obtained 
by integrating the spectral energy distribution from the optical/UV to the X-rays at the time of peak brightness. 
However, in practice, we cannot observe in the extreme UV (EUV) because of Galactic extinction. Therefore, an extrapolation from 
a single or several wave bands to obtain bolometric luminosity is required. Different approaches have been followed 
in the literature.  Some only measure the luminosity in the observed band without any further extrapolation. The
emission of an accretion disk with an inner edge at the ISCO, as in AGNs or BH X-ray binaries, is broader than a single 
blackbody. As in AGNs, some authors apply a bolometric correction, which can be up to a factor of 10, to account 
for an unobservable EUV bump.  However, an elliptical accretion disk is truncated at an inner edge much larger than 
the ISCO: $a_{\rm in} \simeq r_{\rm ms} /(1-e_{\rm d})\sim r_{\rm ms} a_{\rm mb} /[(1-e_{\rm mb}) a_{\rm mb}] 
\sim [r_{\rm ms}/r_{\rm ISCO}]  [a_{\rm mb} /r_{\rm p}] r_{\rm ISCO} \sim 50 \beta m_*^{-1/3} M_6^{1/3} r_{\rm 
ISCO}$. Hence, the disk emission in the EUV and soft X-ray wave bands is expected to be much less significant than 
that of a standard thin accretion disk. The observed spectral energy distributions of TDEs can be fit well by a single 
blackbody with a temperature of about $10^4 \,{\rm K}$, much lower than the prediction of a standard thin accretion 
disk with a typical temperature of $T \gtrsim 10^5\,{\rm K}$ \citep[e.g.,][]{Gezari2012,Holoien2014,Holoien2016a,
Brown2017}. Therefore, many authors approximate the observed spectral energy distribution by a single blackbody 
and then determine the bolometric luminosity by integrating over this single blackbody. Our paper follows the latter 
method.  We calculate the bolometric luminosity by integrating over a single blackbody for the optical and UV 
radiation and then add the contribution from the soft X-ray band at the time of peak brightness.  The only exception 
is for the X-ray TDE XMMSL1 J0740, for which we use the bolometric luminosity from \citet{Saxton2017}. The optical/UV
fluxes are corrected for Galactic extinction and host galaxy starlight. No correction for internal extinction is made 
because there is no significant evidence of internal dust extinction reported for most TDEs. Based on the ratio of 
\ion{He}{2} $\lambda 3203$/$\lambda 4686$, \citet{Gezari2012} suggest that internal extinction might be important
for PS1-10jh, but the origin of the broad optical lines remains unclear, likely arising from an optically thick outflow 
envelope \citep{Roth2016} or a highly eccentric, optically thick accretion disk \citep{Liu2017}.  For GALEX D23H-1, 
significant extinction from the host galaxy can be deduced from the Balmer decrement, but the extinction along 
the line of sight to the flare may be different \citep{Gezari2009}. Except for the TDE candidate SDSS J0952+2143, 
we obtain optical luminosities for most optical/UV TDEs by integrating the blackbody fit to the spectral energy 
distributions, whose temperatures are obtained from multiwave band observations at the peak of the light curves. 
In the event that temperature at the peak is unavailable, we extrapolate it from observations after the peak 
assuming a constant temperature. For SDSS J0952+2143, only unfiltered observations are available at the time of 
peak brightness, and we approximate the optical-UV spectral energy distribution with a blackbody of temperature
$3.5\times10^4\, {\rm K}$  \citep{Komossa2008b}.
 
When available, the X-ray luminosity at the peak brightness is included in the bolometric luminosity. However, except for ASASSN-14li \citep{Brown2017}, the X-ray radiation at peak brightness is either undetected or insignificant for 
most of the optical/UV TDEs.  The optically selected TDE candidate GALEX D1-9 was detected in X-rays $\sim$2.1 yr after the peak \citep{Gezari2008}, but its contribution at peak brightness is unknown.  Its behavior may be 
similar to that of other optical TDEs that were observed to be extremely weak in the X-rays at the time of the peak 
but subsequently became much stronger at late times (ASASSN-15oi: \citealt{Gezari2017}; PTF09axc, PTF09ge, 
and ASASSN-14ae: \citealt{Jonker2020}).  The peak bolometric luminosity of GALEX D1-9 (Table~\ref{tab:obs}) is 
the integral of the blackbody fit to the optical/UV spectral energy distribution at the peak brightness. Because it 
is difficult to estimate the uncertainties of the peak bolometric luminosities, we assign to them an uncertainty of 
0.2~dex ($\sim$60\%).

We note that the total radiation energies after peak for some sources are integrated only up to the end of their
observational campaign. To correct for this limitation, we calculate the total radiation energy $\Delta{E}$ by extrapolating the observations to infinite time with Equation~(\ref{eq:accr}), using 
\begin{equation}
\Delta E= {\Delta E_0 \over 1-\left[L_{\rm e}/L_{\rm p}\right]^{1-1/n}} 
\label{eq:etot}
\end{equation}
with $n=5/3$, where $\Delta{E}_0$ is the radiation energy integrated from the peak time 
$t_{\rm p}$ to time $t_0$ of the end of the observational campaign and $L_{\rm e}$ is the luminosity at $t_0$. 
Two objects required special treatment. For ASASSN-14ae, the bolometric luminosity is best fit with an exponential \citep{Holoien2014}, and for PS18kh, which rebrightened at $\sim$50 days (rest frame) after the peak until 70 days after the peak when the observational campaign ended, we extrapolate the observations by assuming a power-law decay of the luminosity given by Equation~(\ref{eq:accr}) and taking $t_{\rm d}$ from \citet{Holoien2019a}.  It is difficult to estimate the uncertainty of the total radiation energy, but for simplicity, we assume it to be 0.2 dex.

\subsection{Consistency between model predictions and observations}
\label{sec:LE-compare}

Figure~\ref{fig:LE} shows the observed peak bolometric luminosity and the total radiation energy as a function of the 
BH mass calculated from the $M_{\rm BH}$--$\sigma_*$ relation. As explained in Section~4.1, we assume that $L_{\rm 
p}$ and $\Delta E$ gave uncertainties of 0.2~dex. Both the observed peak bolometric luminosity and the total 
radiation energy correlate tentatively with the BH mass, consistent with the results obtained by \citet{Wevers2017,Wevers2019}.  These results, in combination with the absence of a connection between blackbody 
temperature and BH mass \citep{Wevers2017,Wevers2019}, are at odds with the predictions of the shock-powered 
model of \citet{Piran2015} but are expected for the elliptical accretion disk model of roughly uniform eccentricity  
\citep{Liu2020}. Figure~\ref{fig:LE} shows that the observed peak bolometric luminosity and the total radiation energy 
are consistent with the expectations from the elliptical accretion disk model with orbital penetration factor $\beta =1$, 
but much lower than those expected with the canonical radiation efficiency $\eta=0.1$. The results suggest that the 
sample TDEs probably result from the tidal disruption of stars of type A or later by supermassive BHs of mass between 
$10^5 \,M_\odot$ and $10^7 \,M_\odot$.  This is consistent with the observation that the host galaxies of most TDEs are 
post-starbursts, for which star formation occurred about a billion years ago, and hence presently have a deficit of B- and O-
type stars but are rich in stars of type A and later \citep{Arcavi2014,French2016, Law-Smith2017a, Graur2018}. The sole 
exception is the star-forming galaxy  SDSS J0952+2143 \citep{Palaversa2016}, whose TDE probably arose from a 
disrupted early A-type or late B-type star. To summarize: given $L_{\rm p}$, $\Delta{E}$, and $M_{\rm BH}$, we can
solve Equations~(\ref{eq:plum}) and (\ref{eq:toteng}) to obtain the mass $M_*$ and orbital penetration factor 
$\beta$.

\subsection{Mass of the star and the accreted fraction}
\label{sec:starmass}

\begin{figure}
\begin{center}
\includegraphics[width=0.9\columnwidth]{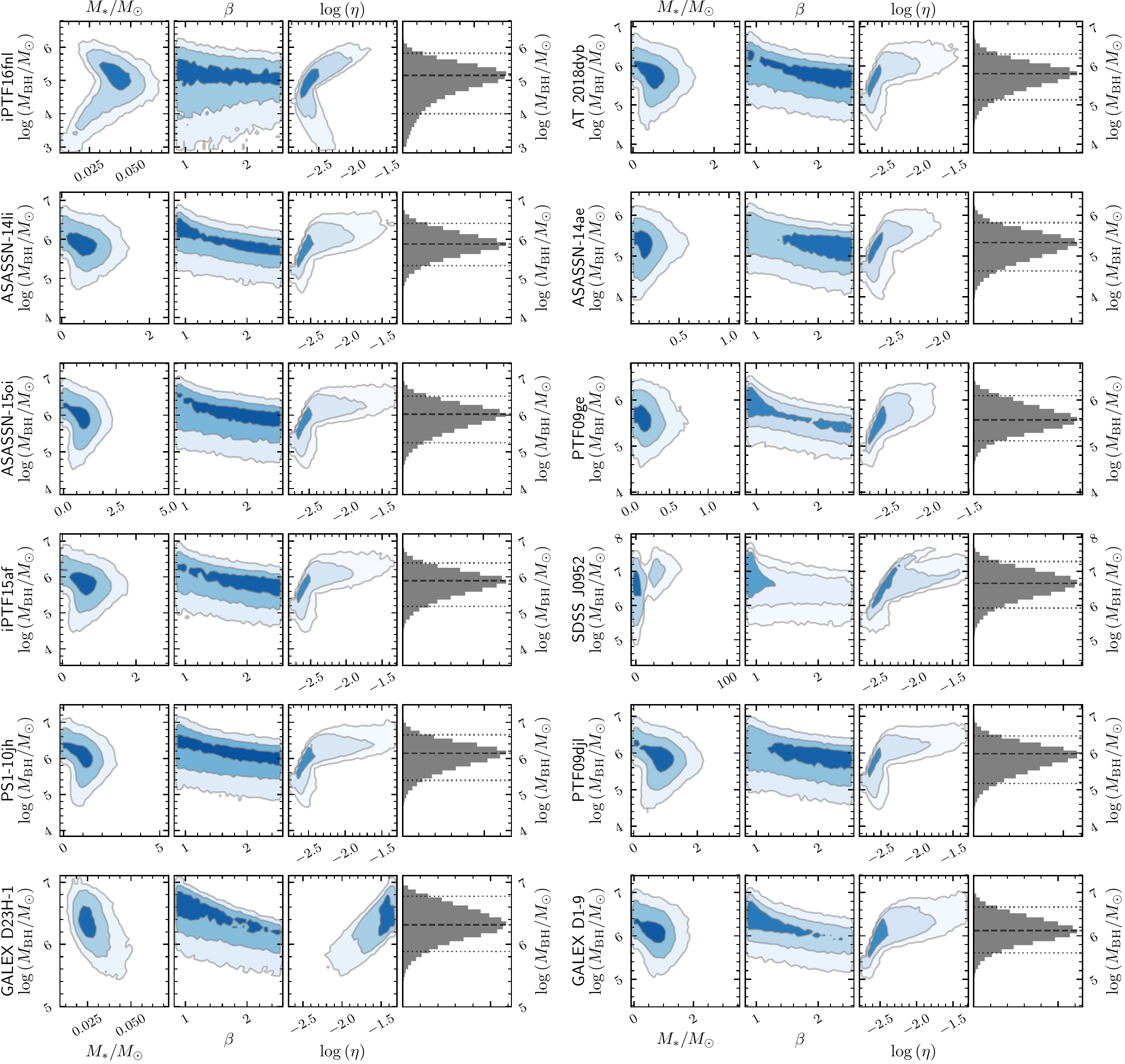}
\caption{Posterior distributions of the model parameters ($M_{\rm BH}$, $M_*$, and $\beta$) 
and the radiation efficiency ($\eta$). The contours are for 1, 2, and 3 $\sigma$. In the histogram 
of $M_{\rm BH}$, the dashed line indicates the BH mass at the peak of the distribution, and the two dotted lines give 
the BH mass ranges at the 90\% confidence level. 
\label{fig:contour_mbh}
}
\end{center}
\end{figure}

With the observations of the BH mass $M_{\rm BH}$, the peak bolometric luminosity $L_{\rm p}$, and the total 
radiation energy after peak $\Delta{E}$ (or apparent accreted stellar mass $\Delta{M}_{\rm app}$), we 
can solve Equations~(\ref{eq:plum}) and (\ref{eq:toteng}) (or (\ref{eq:accmass})) to obtain the mass 
and orbital penetration factor of the star. We solve the equations using Markov Chain Monte Carlo (MCMC) with 
the python package emcee (\citealt{Foreman-Mackey2013}). The likelihood function is 
\begin{equation}
-{1\over2} \left[{(\log{L_{\rm p}} - \log{L_{\rm p}^\prime})^2 \over \sigma_{\rm L}^2} + \ln(2\pi
    \sigma_{\rm L}^2) + {(\log{\Delta{E}} - \log{\Delta{E}^\prime})^2 \over \sigma_{\rm E}^2} + \ln(2\pi
    \sigma_{\rm E}^2)\right] ,
    \label{eq:likelihood}
\end{equation}
where $L_{\rm p}$ and $\Delta{E}$ are the observed peak bolometric luminosity and the 
total radiation energy after peak, respectively, and $L_{\rm p}^\prime$ and $\Delta{E}^\prime$ are the
estimates of Equations~(\ref{eq:plum}) and (\ref{eq:toteng}), respectively, with the input parameters ($M_{\rm BH}$, 
$M_*$, and $\beta$), $\sigma_{\rm L}$ of 0.2 dex is the uncertainty of the  peak bolometric luminosity, 
and $\sigma_{\rm E}$ of 0.2 dex is the uncertainty of the total radiation energy after peak.
The prior parameters of the MCMC experiments are the BH mass $M_{\rm BH}$, the stellar 
mass $M_*$, and the orbital penetration factor $\beta$ of star. The prior distribution of $M_*$ and $\beta$ are 
uniform in the ranges of $0.01\,M_\odot<M_*<150\,M_\odot$ and $0.9\leqslant \beta \leqslant 2.5$, respectively. 
By fitting the multiwavelength light curves of a sample of TDEs, \citet{Mockler2019} showed that most of the 
TDEs have $\beta\simeq 1$ with a range $0.9\la \beta\la 1.8$. We adopted the lower limit $\beta_{\rm l} = 
0.9$ because the method would give a poor constraint on $\beta$ and the survey of TDEs would prefer detecting 
the full tidal disruption of stars to the partial disruptions as the former would give rise to higher peak luminosity 
and longer duration of TDE flares. Although an upper limit $\beta_{\rm u}=2.5$ is small and numerical hydrodynamic 
simulations of tidal disruptions with larger penetration factor $\beta$ have been carried out in the literature
\citep{Evans2015, Sadowski2016, Darbha2019}, we adopt $\beta_{\rm u}=2.5$ in this work and expect that 
the results except for $\beta$ are not changed significantly by increasing the upper limit of the  penetration factor. 
The reason is as follows. The posterior distributions of $\beta$ in Figure~\ref{fig:contour_mbh} (also in
Figure~\ref{fig:mcmc-contour} and Figure~\ref{fig:mcmc-contour2}) show that $\beta$ is not constrained 
well for the sample sources. Both the peak bolometric luminosity $L_{\rm p}$ and the total radiation energy 
after peak $\Delta{E}$ depend on the penetration factor $\beta$ mostly because of the parameter $\Delta$ 
of the radiation efficiency $\eta$. Equations~(\ref{eq:plum}) and (\ref{eq:toteng}) show that for $\Delta \lesssim 
1$, $L_{\rm p}$ and $\Delta E$ depend very weakly on $\beta$ and their solutions would give poor constraints on  
the penetration factor. For $\Delta \gg 1$ (or $\beta \gg 1$) or $\Delta \ll 1$ (or $\beta \ll 1$), both $L_{\rm p}$ and $\Delta{E}$ change significantly with $\beta$. The  penetration factor $\beta$ can be well determined, 
and Equations~(\ref{eq:plum}) and (\ref{eq:toteng}) should be solved with the results of numerical hydrodynamic 
simulations with much larger ranges of penetration factor \citep{Evans2015, Sadowski2016, Darbha2019}. Because 
the 18 sample sources have  $\Delta \lesssim 1$, the adopted range of penetration factor $0.9 \leqslant \beta \leqslant 
2.5$ is reasonable, and the obtained penetration factors of the sample sources including those with $\beta\sim 
2.5$ are rough estimates with large uncertainties.  The prior distribution 
of the parameter $M_{\rm BH}$ is a normal distribution whose mean and variance corresponding to the observed values and their uncertainties, respectively, given in Column 7 of Table~\ref{tab:obs}. The MCMC chain includes 100 
walkers with each walker consisting of $10^4$ steps. The first 50\% of the steps of each walker are removed for 
burn-in and one set of the parameters is saved every five steps for the rest of the walkers. For each walker, the parameters 
begin with the local best-fit results from the least-squares method plus a small random offset.

Figure~\ref{fig:contour_mbh} shows the posterior distributions of model parameters ($M_{\rm BH}$,
$M_*$, and $\beta$) of the MCMC experiments, and Table~\ref{tab:mcmc_mbh} gives the results of the
parameters $M_*$ and $\beta$ and the associated uncertainties at the 90\% confidence level 
obtained with the MCMC method. Figure~\ref{fig:contour_mbh} shows that the BH and stellar masses 
of TDEs can be well determined, but the orbital penetration factor of the star is constrained poorly.. These results are consistent with the arguments for TDEs with $\Delta =0.25 \beta^{3} r_*^{-2} m_*^{1/3} M_6^{5/3} \la 1$, or
with $\beta\sim 1$ and $M_{\rm BH} \la 10^{6.5} \,M_\odot$ given at the end of Section\ref{sec:LEdigram}.
Our MCMC experiments show that some TDEs may have two solutions, 
one associated with a main-sequence star and the other associated with a BD. This is possible
because the main-sequence stars and BDs have significantly different relations of stellar mass 
and radius, with a dramatic transition from $M_*\sim 0.07$ to $0.08 \,M_\odot$: The radius increases with 
mass for main-sequence stars but decreases with mass for a BD. We do not simply remove the 
solutions associated with BDs. We compare the probabilities of the posterior distributions 
of the two solutions and adopt the one with higher posterior probability to be the main solution of the TDE. 
Figure~\ref{fig:contour_mbh} gives the posterior distributions of the model parameters 
of the main solution. We give our conclusions based on the primary solutions of the TDEs. However, when 
the ratio of the probabilities associated with the two solutions is lower than $3:1$, we keep both solutions and give the secondary solution in the row after the primary of 
Table~\ref{tab:mcmc_mbh}.

\begin{figure}
\begin{center}
\includegraphics[width=0.9\columnwidth]{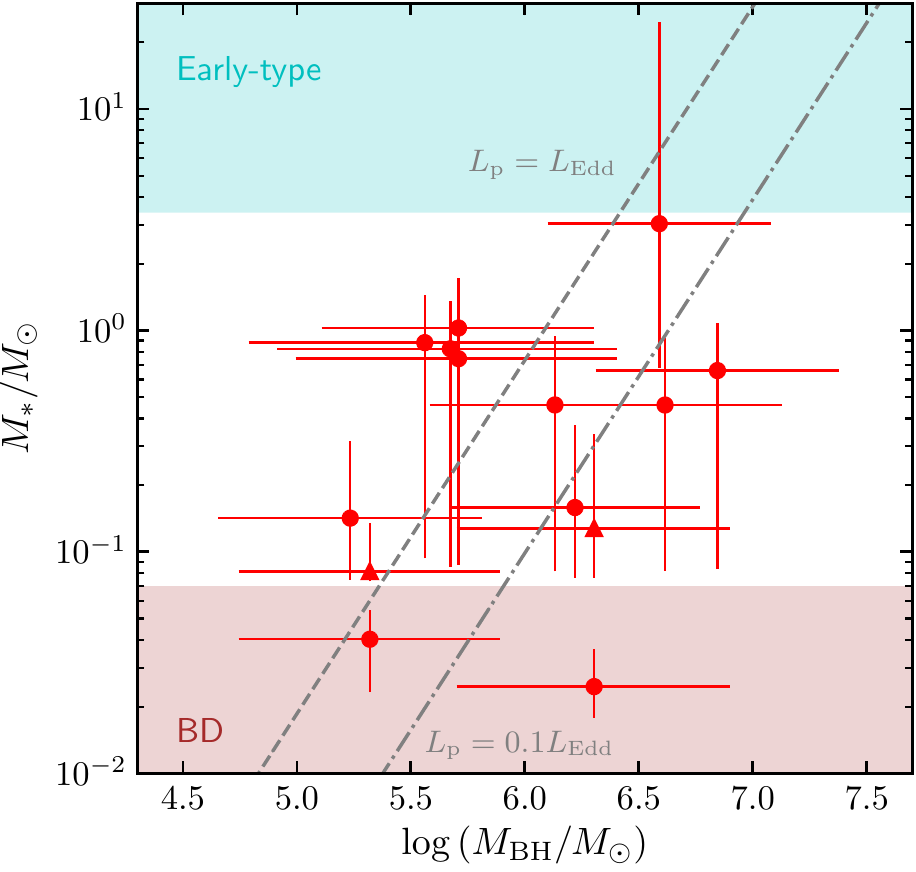}
\caption{The stellar mass $M_*$ (filled circles) vs. BH mass. The two color-shaded regions are for early-type stars (Early-type) and brown-dwarfs (BD). The filled triangles are for the secondary
solutions of iPTF16fnl and GALEXD23H-1. The stellar mass and the associated uncertainty at 90\% 
confidence level are calculated with the MCMC method. The dashed and dash--dotted lines refer to the observational selection effect according to Equation~(\ref{eq:seleffect_Edd}) for $L_{\rm p} = L_{\rm Edd}$ and $0.1\, L_{\rm Edd}$, respectively (see 
Section~\ref{sec:starmass} for details).
\label{fig:mstar}}
\end{center}
\end{figure}

Figure~\ref{fig:mstar} shows the masses of the stars of the 12 TDEs obtained with the MCMC experiments, 
including both the primary and secondary possible solutions. Figure~\ref{fig:mstar} 
and Table~\ref{tab:mcmc_mbh} show that the stars have spectral types ranging from A- through 
M-type main-sequence stars to BDs. Among all the 12 sample sources 
with BH mass obtained with $M_{\rm BH}$--$\sigma_*$ relations, iPTF16fnl and GALEX D23H-1 both have two solutions, with the primary solutions associated with a BD and the secondary associated with a main-sequence 
star. TDE iPTF16fnl has the lowest total radiation energy and the second lowest peak bolometric 
luminosity after GALEX D23H-1. It also has a light curve of the decay timescale that is among the shortest and has
a BH among the lowest mass \citep{Blagorodnova2017,Onori2019}. Therefore, the main solution for iPTF16fnl 
associated with a BD is likely the real solution of the source. TDE GALEX D23H-1 has 
the lowest peak bolometric luminosity and is one of the sources with the lowest total radiation energy. These factors lead to the solution of a BD. However, the low peak bolometric luminosity and total radiation 
energy of GALEX D23H-1 might not be intrinsic but due to a possible intrinsic dust extinction of the 
host galaxy because a global extinction has been detected with the Balmer decrement of the 
\ion{H}{2} regions and the extinction in the line of sight to the flare might be important \citep{Gezari2009}. Except for iPTF16fnl and GALEX D23H-1,  the other 10 sample sources with observations of a BH mass have solutions 
associated with late-type main-sequence or A-type stars. Our result that most stars of the TDE sources are
late-type main-sequence stars or BDs is well consistent with the fact that 
the host galaxies of most TDEs except for SDSS J0952+2143 are post-merger E+A galaxies, with the last burst occurring about a billion years ago, so that the stellar population in the centers are dominated by stellar types A and later \citep{Arcavi2014,French2016,French2017}. 

When we calculated the peak bolometric luminosity with $L_{\rm p}= \eta \dot{M}
_{\rm p} c^2$ and the total radiation energy $\Delta{E} = \eta \Delta{M_*} c^2$ with the radiation efficiency 
$\eta$ given by Equation~(\ref{eq:teff}), we have implicitly assumed that the peak bolometric luminosity is 
sub-Eddington. However, when the peak fallback rate $\dot{M}_{\rm p}$ is near or above the Eddington accretion
rate $\dot{M}_{\rm Edd} = L_{\rm Edd} / (\eta c^2)$, the peak luminosity $L_{\rm p}$ scales as \citep{Paczynski1980}
$L_{\rm p} \simeq L_{\rm Edd} \left[1 + \log\left(\dot{M}_{\rm p} /\dot{M}_{\rm Edd}\right)\right]$ because of the 
photon 
trapping and because a significant fraction of radiation is advected onto the BH \citep{Abramowicz1988}. For TDEs
with $\dot{M}_{\rm p} \ga \dot{M}_{\rm Edd}$, the light curves are capped at the 
Eddington luminosity. To compare the observations of TDEs with the predictions of the elliptical accretion disk model,
we excluded from the sample TDE sources the candidates with an extended plateau in their light curves (most of these TDE candidates are in AGNs). Meanwhile, transient surveys are more likely 
to detect bright TDEs. They preferentially detect TDEs with a peak luminosity at or near the Eddington 
luminosity. Therefore, if we assume $L_{\rm p}\sim L_{\rm Edd}$, Equation~(\ref{eq:plum}) gives
\begin{eqnarray}
m_* &\simeq & \left({1.25\over 1.78}\right)^{6/(5+9\zeta)} \left[\left({\eta_0\over 0.059}\right)^{-1}\left({A_\gamma 
         \over 1.328}\right)^{-1} \beta\left(1+\Delta\right)^{-1}\right]^{6/(5+9\zeta)} \left({L_{\rm p}\over L_{\rm Edd}}
          \right)^{6/(5+9\zeta)} M_6^{11/(5+9\zeta)} \cr
        &\simeq & 0.70^{6/(5+9\zeta)}  \left({L_{\rm p}\over L_{\rm Edd}}\right)^{6/(5+9\zeta)} 
        M_6^{11/(5+9\zeta)} .
\label{eq:seleffect}
\end{eqnarray}
For $\zeta \simeq 0.21$, we have 
\begin{equation}
m_* \simeq 0.70^{6/6.89} \left({L_{\rm p}\over L_{\rm Edd}}\right)^{6/6.89} M_6^{11/6.89} .
\label{eq:seleffect_Edd}
\end{equation}
Figure~\ref{fig:mstar} also shows the selection effect according to Equation~(\ref{eq:seleffect_Edd}) for $L_{\rm p} = L_{\rm 
Edd}$ and $L_{\rm p} =0.1 L_{\rm Edd}$. Taking into account the large intrinsic scatter of the $M_{\rm BH}$--$\sigma_*$ relation ($\sigma=0.49$~dex or $3\sigma=1.47$~dex), Figure~\ref{fig:mstar} shows that the mass of the 
star may correlate with the mass of BH, which is consistent with the suggestion of Equation~(\ref{eq:seleffect_Edd}).

Given $M_{\rm BH}$, $M_*$, and $\beta$, we can calculate the radiation efficiency,
$\eta=\eta(M_{\rm BH},R_*,M_*,\beta)$, with Equation~(\ref{eq:teff}) and the MCMC method. The posterior
distributions of $\eta$ are given
in Figure~\ref{fig:contour_mbh}, and the radiation efficiencies and the associated uncertainties at the 90\% 
confidence level are given in Table~\ref{tab:mcmc_mbh}. 
These results are also shown in Figure~\ref{fig:eff}, where the BH masses are computed from the $M_{\rm BH}$--$\sigma_*$ relation. 
Figure~\ref{fig:eff} shows that the radiation efficiencies are much lower than the 
canonical value $\eta=0.1$ that is commonly adopted in the literature. All the TDE sources except 
GALEX D23H-1 have a typical radiation efficiency $\log(\eta) \simeq -2.57$ or $\eta \simeq 2.7\times 10^{-3}$, 
which is about $37$ times lower than the canonical radiation efficiency. 
A low radiation efficiency would lead to a low peak bolometric luminosity and a low total 
radiation energy for a given accretion rate of matter. In other words, given the observed peak bolometric luminosity and the total radiation energy after peak, we would obtain a much higher apparent
accretion rate and total accreted stellar material due to the low radiation efficiency. Our result suggests that the low bolometric peak luminosity and
total radiation energy of TDEs result from the low conversion efficiency of matter into radiation associated with the elliptical
accretion disk.

\begin{deluxetable*}{cccccc}
\tablecaption{Results of the MCMC experiments for the TDEs with BH mass provided
\label{tab:mcmc_mbh}}
\tabletypesize{\small}
\tablewidth{0pt}
\tablehead{
\colhead{Name} & \colhead{$M_*$} & \colhead{$\beta$} & 
\colhead{$\log (\eta)$} & \colhead{$\Delta M_*$} & 
\colhead{$\Delta M_*/M_*$} \\
\colhead{} & \colhead{($M_\odot$)} & \colhead{} & 
\colhead{} & \colhead{($M_\odot$)} & 
\colhead{}
}
\colnumbers
\startdata
iPTF16fnl\tablenotemark{\scriptsize{a}}	  & $0.040^{+0.015}_{-0.017}$ & $0.9^{+1.4}_{-0.0}$ & $-2.68^{+0.45}_{-0.12}$ &$0.012^{+0.005}_{-0.005}$ & $0.34^{+0.00}_{-0.09}$ \\ 
& $0.081^{+0.054}_{-0.008}$ & $2.5^{+0.0}_{-1.3}$ & $-2.74^{+0.18}_{-0.04}$ & $0.026^{+0.011}_{-0.008}$ & $0.34^{+0.00}_{-0.10}$ \\ 
AT 2018dyb	& $0.46^{+0.47 }_{-0.38}$ & $2.5^{+0.0}_{-1.4}$ & $-2.64^{+0.53}_{-0.08}$ & $0.13^{+0.13}_{-0.11}$ & $0.34^{+0.00}_{-0.09}$ \\ 
ASASSN-14li	  & $0.46^{+0.49 }_{-0.38}$ & $0.9^{+1.4}_{-0.0}$ & $-2.57^{+0.54}_{-0.14}$ & $0.15^{+0.13}_{-0.13}$ & $0.34^{+0.00}_{-0.09}$ \\ 
ASASSN-14ae	  & $0.14^{+0.17 }_{-0.07}$ & $2.5^{+0.0}_{-1.4}$ & $-2.69^{+0.30}_{-0.06}$ & $0.043^{+0.047}_{-0.022}$ & $0.34^{+0.00}_{-0.09}$ \\ 
ASASSN-15oi	  & $0.88^{+0.56 }_{-0.79}$ & $0.9^{+1.4}_{-0.0}$ & $-2.61^{+0.65}_{-0.09}$ & $0.25^{+0.17}_{-0.23}$ & $0.34^{+0.00}_{-0.09}$ \\ 
PTF09ge	  & $0.16^{+0.21 }_{-0.08}$ & $0.9^{+1.4}_{-0.0}$ & $-2.59^{+0.35}_{-0.14}$ & $0.050^{+0.060}_{-0.028}$ & $0.34^{+0.00}_{-0.09}$ \\ 
iPTF15af	  & $0.66^{+0.42 }_{-0.57}$ & $2.4^{+0.0}_{-1.4}$ & $-2.62^{+0.56}_{-0.08}$ & $0.17^{+0.14}_{-0.15}$ & $0.34^{+0.00}_{-0.09}$ \\ 
SDSS J0952+2143  & $3.03^{+21.55}_{-2.35}$ & $0.9^{+1.3}_{-0.0}$ & $-2.30^{+0.53}_{-0.28}$ & $0.98^{+0.71}_{-0.76}$ & $0.31^{+0.01}_{-0.28}$ \\ 
PS1-10jh	  & $1.02^{+0.69 }_{-0.89}$ & $0.9^{+1.4}_{-0.0}$ & $-2.55^{+0.68}_{-0.13}$ & $0.32^{+0.18}_{-0.29}$ & $0.34^{+0.00}_{-0.09}$ \\ 
PTF09djl	  & $0.83^{+0.53 }_{-0.74}$ & $2.4^{+0.1}_{-1.4}$ & $-2.61^{+0.58}_{-0.10}$ & $0.24^{+0.15}_{-0.21}$ & $0.34^{+0.00}_{-0.09}$ \\ 
GALEX D23H-1\tablenotemark{\scriptsize{a}}  & $0.025^{+0.012}_{-0.007}$ & $0.9^{+1.4}_{-0.0}$ & $-1.40^{+0.04}_{-0.41}$ & $0.0075^{+0.0039}_{-0.0025}$ & $0.34^{+0.00}_{-0.09}$ \\ 
& $0.13^{+0.21 }_{-0.05}$ & $0.9^{+1.3}_{-0.0}$ & $-2.57^{+0.37}_{-0.13}$ & $0.042^{+0.059}_{-0.020}$ & $0.34^{+0.00}_{-0.08}$ \\ 
GALEX D1-9	  & $0.74^{+0.44 }_{-0.66}$ & $0.9^{+1.4}_{-0.0}$ & $-2.54^{+0.68}_{-0.14}$ & $0.23^{+0.13}_{-0.20}$ & $0.34^{+0.00}_{-0.09}$ \\ 
\enddata
\tablecomments{The BH masses are calculated with the $M_{\rm BH}$--$\sigma_*$ relation. The 
uncertainties of the model parameters are at the 90\% confidence level obtained with MCMC 
experiments. }
\tablenotetext{a}{The source has two possible solutions, and the main one with the higher probability is given at the first entry.}
\end{deluxetable*}

%

\begin{figure}
\begin{center}
\includegraphics[width=0.9\columnwidth]{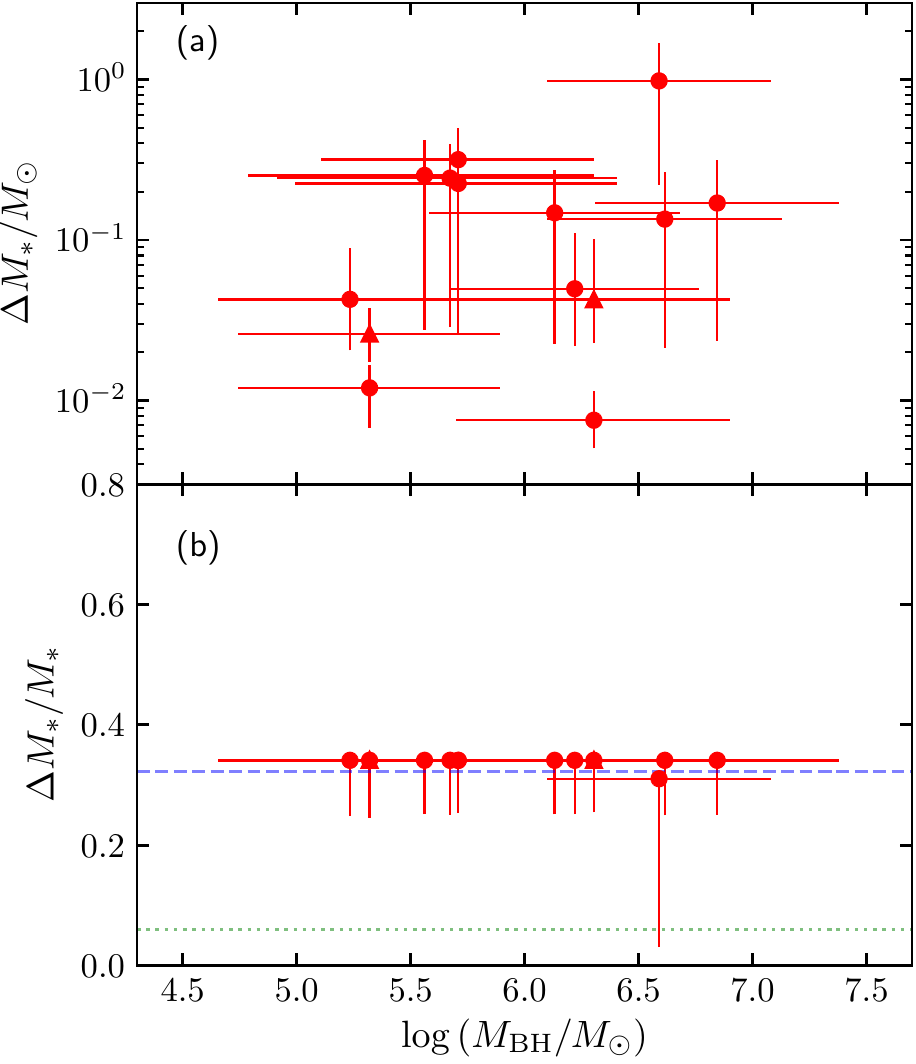}
\caption{Accreted stellar mass after peak (a) and the relative accreted fraction (b) vs.
BH mass for the sample sources in Table~\ref{tab:mcmc_mbh}. The dashed and dotted lines in panel (b) show the prediction of the polytropic model assuming $\gamma=5/3$ for low-mass stars and $\gamma=4/3$ 
for high-mass stars,
respectively. In both cases, $\beta=1$ is assumed. The result of SDSS J0952+2143 with $M_*\simeq 3 \,M_\odot$ is obtained with 
the hybrid polytropic model and close to the predictions with $\gamma=5/3$.
\label{fig:macc}}
\end{center}
\end{figure}

Table~\ref{tab:mcmc_mbh} and Figure~\ref{fig:macc} give the masses of the accreted material $\Delta{M}_*$ 
and the fractions with respect to the masses of the disrupted stars (i.e., the accreted fraction of stellar mass) $\Delta{M}_*/M_*$ derived from the MCMC 
experiments. The accreted stellar mass together with the conversion efficiency gives the expected total radiation 
energy $\Delta{E}^\prime=\eta\Delta{M}_*c^2$ of Equation~(\ref{eq:likelihood}). In Table~\ref{tab:mcmc_mbh} 
and Figure~\ref{fig:macc}, we also give the associated uncertainties of $\Delta{M}_*$ and $\Delta{M}_*/M_*$ at 
the 90\% confidence level. Equation~(\ref{eq:accm}) shows that 
the relative accreted stellar mass can be obtained with $\Delta{M}_*/M_* \simeq (1-n)^{-1} A_\gamma B_\gamma 
\simeq 1.5 A_\gamma B_\gamma$, which depends on the stellar structure and orbital penetration factor $\beta$ 
\citep{Lodato2009,Guillochon2013,Golightly2019a,Law-Smith2019,Ryu2019a}. Therefore, in Figure~\ref{fig:macc}, we 
also show the $\Delta{M}_*/M_*$ calculated with the 
empirical formulae of $A_\gamma$ and $B_\gamma$ in the appendix of \citet{Guillochon2013} for both $\gamma=5/3$ and $4/3$, while fixing $\beta=1$ . 
It shows that the total accreted material after peak 
significantly varies from about $10^{-2} \,M_\odot$ of GALEX D23H-1 and iPTF16fnl to about $1 \,M_\odot$ of SDSS 
J0952+2143, but the accreted material relative to the total mass of star of our TDEs except SDSS J0952+2143 
is approximately constant, with $\Delta{M}_*/M_* \sim 0.34$, which is given by the hydrodynamic 
simulations of the tidal disruption of low-mass star with polytropic index $\gamma=5/3$.  For SDSS J0952+2143, the star 
has a mass of about $3.03 \,M_\odot$ and is described with our hybrid model. The relative accreted stellar mass 
of SDSS J0952+2143 is close to the expectation of the polytropic model $\gamma=5/3$, but with very large 
uncertainties.

\section{Weighing BHs using TDEs}
\label{sec:bhmass}

\subsection{Deriving the BH and stellar masses with $L_{\rm p}$ and $\Delta{E}$}
\label{sec:method} 

Since a massive BH could be a member of a supermassive BH binary, might lie in a globular cluster, or have an off-nuclear 
position, it is important to have an alternative method other than the $M_{\rm BH}$--$\sigma_*$ relation to calculate 
the mass of the BH. Equations~(\ref{eq:pmass}) and (\ref{eq:accm}) show that provided the peak accretion 
rate $\dot{M}_{\rm p}$ and the total accreted material $\Delta{M}_*$, one could uniquely determine 
the masses of the BH and the star by solving these two equations. However, we cannot directly measure 
$\dot{M}_{\rm p}$ and $\Delta{M}_*$ but the peak bolometric luminosity $L_{\rm p}$ and the total radiation energy 
$\Delta{E}$, which depend not only on the masses of the BH and the star, but also on the radiation efficiency, 
the latter of which depends on the orbital penetration factor $\beta$. The solutions of the stellar mass and the BH 
mass become functions of the penetration factor and would be expected to be determined observationally with 
larger uncertainties. Figure~\ref{fig:contour_mbh} shows that even though we have the measurement of BH 
mass with the $M_{\rm BH}$--$\sigma_*$ relation, the uncertainty in $\beta$ is as large as the entire range of the 
prior. The large uncertainty is consistent with the arguments in Section~\ref{sec:LEdigram} that the peak luminosity 
and the total radiation energy are nearly independent of the penetration factor for the range $0.7 \la \beta \la 2.1$ 
and implies that the mass of the BHs do not significantly couple with the penetration factor. We expect to determine
the masses of the BHs and the stars with small uncertainties by solving Equations~(\ref{eq:plum}) and (\ref{eq:toteng}) 
given the observed $L_{\rm p}$ and $\Delta{E}$.  The uncertainty in $\beta$ should not result in a large uncertainty 
in the measurement of the masses of the BHs and the stars.

With the observations of $L_{\rm p}$ and $\Delta{E}$, we solve the equations with the MCMC method 
as described in Section~\ref{sec:starmass}, except that the prior distributions of all the three parameters 
$M_{\rm BH}$, $M_*$ and $\beta$ are now uniform in the ranges $10^3 \,M_\odot \leqslant M_{\rm BH} \leqslant 
10^9 \,M_\odot$, $0.01\,M_\odot <M_* <150 \,M_\odot$, and $0.9\leqslant \beta \leqslant 2.5$. The large ranges 
for the masses of BHs and stars require a large amount of computational time. To enhance the convergence
rate of the MCMC experiments, we start the experiments 
with $\beta=1$, and with the masses of the BH and the star that are calculated from Equations~(\ref{eq:plum}) and 
(\ref{eq:toteng}) and the observed $L_{\rm p}$ and $\Delta{E}$ for $\beta=1$. We use these initial conditions because TDEs are expected to predominantly occur at $\beta
\sim 1$ \citep{Kochanek2016,Stone2016}. Since the results of the masses of the BH and the star depend only 
weakly on the penetration factor, the solutions obtained with $\beta=1$ are good approximations.

\begin{deluxetable*}{ccccccc}
\tablecaption{Results of the MCMC experiments for all the sample sources
\label{tab:mcmc}}
\tabletypesize{\small}
\tablewidth{0pt}
\tablehead{
\colhead{Name} & \colhead{${\rm \log} (M_{\rm BH})$} & \colhead{$M_*$} & \colhead{$\beta$} & 
\colhead{${\rm \log} (\eta)$} & \colhead{$\Delta M_*$} & 
\colhead{$\Delta M_*/M_*$} \\
\colhead{} & \colhead{($M_\odot$)} & \colhead{($M_\odot$)} & \colhead{} & 
\colhead{} & \colhead{($M_\odot$)} & 
\colhead{}
}
\colnumbers
\startdata
iPTF16fnl\tablenotemark{\scriptsize{a}}	 & $5.16^{+0.66}_{-1.16}$ & $0.040^{+0.015 }_{-0.017}$ & $0.9^{+1.4}_{-0.0}$ & $-2.68^{+0.43}_{-0.13}$ & $0.012^{+0.005}_{-0.005}$ & $0.34^{+0.00}_{-0.09}$ \\ 
& $4.96^{+0.51}_{-0.45}$ & $0.082^{+0.055}_{-0.008}$ & $2.5^{+0.0}_{-1.3}$ & $-2.73^{+0.17}_{-0.05}$ & $0.025^{+0.013}_{-0.007}$ & $0.34^{+0.00}_{-0.10}$ \\ 
AT 2018dyb	 & $5.80^{+0.50}_{-0.67}$ & $0.46^{+0.48}_{-0.38}$ & $2.5^{+0.0}_{-1.4}$ & $-2.63^{+0.52}_{-0.09}$ & $0.14^{+0.13}_{-0.12}$ & $0.34^{+0.00}_{-0.09}$ \\ 
ASASSN-14li	 & $5.88^{+0.53}_{-0.55}$ & $0.41^{+0.53 }_{-0.33}$ & $0.9^{+1.4}_{-0.0}$ & $-2.56^{+0.53}_{-0.14}$ & $0.14^{+0.13}_{-0.12}$ & $0.34^{+0.00}_{-0.09}$ \\ 
ASASSN-14ae	 & $5.32^{+0.49}_{-0.69}$ & $0.14^{+0.18 }_{-0.06}$ & $2.5^{+0.0}_{-1.4}$ & $-2.69^{+0.31}_{-0.06}$ & $0.043^{+0.047}_{-0.022}$ & $0.34^{+0.00}_{-0.09}$ \\ 
ASASSN-15oi	 & $6.03^{+0.49}_{-0.78}$ & $0.89^{+0.53 }_{-0.81}$ & $2.3^{+0.1}_{-1.3}$ & $-2.59^{+0.63}_{-0.10}$ & $0.26^{+0.16}_{-0.23}$ & $0.34^{+0.00}_{-0.09}$ \\ 
PTF09ge	 & $5.57^{+0.53}_{-0.45}$ & $0.15^{+0.22 }_{-0.07}$ & $0.9^{+1.4}_{-0.0}$ & $-2.59^{+0.35}_{-0.14}$ & $0.048^{+0.060}_{-0.026}$ & $0.34^{+0.00}_{-0.09}$ \\ 
iPTF15af	 & $5.89^{+0.50}_{-0.71}$ & $0.57^{+0.51 }_{-0.48}$ & $2.5^{+0.0}_{-1.4}$ & $-2.62^{+0.57}_{-0.08}$ & $0.17^{+0.15}_{-0.14}$ & $0.34^{+0.00}_{-0.09}$ \\ 
SDSS J0952+2143	 & $6.65^{+0.64}_{-0.72}$ & $2.99^{+21.11}_{-2.35}$ & $0.9^{+1.3}_{-0.0}$ & $-2.31^{+0.54}_{-0.27}$ & $0.97^{+0.72}_{-0.75}$ & $0.31^{+0.01}_{-0.28}$ \\ 
PS1-10jh	 & $6.14^{+0.52}_{-0.75}$ & $1.05^{+0.64 }_{-0.95}$ & $0.9^{+1.4}_{-0.0}$ & $-2.57^{+0.70}_{-0.12}$ & $0.30^{+0.20}_{-0.27}$ & $0.34^{+0.00}_{-0.09}$ \\ 
PTF09djl	 & $5.98^{+0.49}_{-0.81}$ & $0.80^{+0.55 }_{-0.71}$ & $2.3^{+0.1}_{-1.3}$ & $-2.61^{+0.61}_{-0.09}$ & $0.24^{+0.16}_{-0.21}$ & $0.34^{+0.00}_{-0.09}$ \\ 
GALEX D23H-1\tablenotemark{\scriptsize{a}} & $6.31^{+0.46}_{-0.43}$ & $0.025^{+0.012 }_{-0.007}$ & $0.9^{+1.4}_{-0.0}$ & $-1.40^{+0.03}_{-0.43}$ & $0.0074^{+0.0040}_{-0.0023}$ & $0.34^{+0.00}_{-0.09}$ \\ 
& $5.64^{+0.55}_{-0.40}$ & $0.12^{+0.22 }_{-0.05}$ & $0.9^{+1.3}_{-0.0}$ & $-2.58^{+0.38}_{-0.13}$ & $0.047^{+0.057}_{-0.024}$ & $0.34^{+0.00}_{-0.08}$ \\
GALEX D1-9	 & $6.12^{+0.54}_{-0.51}$ & $0.71^{+0.47 }_{-0.63}$ & $0.9^{+1.4}_{-0.0}$ & $-2.54^{+0.69}_{-0.14}$ & $0.20^{+0.15}_{-0.18}$ & $0.34^{+0.00}_{-0.09}$ \\ 
\hline
XMMSL1 J0740 & $5.63^{+0.52}_{-0.91}$ & $0.42^{+0.52 }_{-0.32}$ & $2.5^{+0.0}_{-1.4}$ & $-2.65^{+0.34}_{-0.07}$ & $0.13^{+0.14}_{-0.09}$ & $0.34^{+0.00}_{-0.09}$ \\ 
ASASSN-19bt	 & $5.88^{+0.52}_{-0.65}$ & $0.53^{+0.51 }_{-0.44}$ & $0.9^{+1.4}_{-0.0}$ & $-2.59^{+0.54}_{-0.12}$ & $0.16^{+0.14}_{-0.14}$ & $0.34^{+0.00}_{-0.09}$ \\ 
AT 2018fyk	 & $6.27^{+0.52}_{-0.79}$ & $1.39^{+0.88 }_{-1.24}$ & $0.9^{+1.4}_{-0.0}$ & $-2.53^{+0.72}_{-0.13}$ & $0.41^{+0.27}_{-0.36}$ & $0.34^{+0.00}_{-0.09}$ \\ 
PS18kh		 & $5.82^{+0.46}_{-0.63}$ & $0.35^{+0.41 }_{-0.27}$ & $0.9^{+1.4}_{-0.0}$ & $-2.64^{+0.53}_{-0.08}$ & $0.11^{+0.11}_{-0.09}$ & $0.34^{+0.00}_{-0.09}$ \\ 
AT 2017eqx	 & $5.56^{+0.51}_{-0.66}$ & $0.25^{+0.31 }_{-0.18}$ & $2.5^{+0.0}_{-1.4}$ & $-2.67^{+0.43}_{-0.06}$ & $0.079^{+0.081}_{-0.058}$ & $0.34^{+0.00}_{-0.09}$ \\ 
PS1-11af	 & $5.71^{+0.52}_{-0.55}$ & $0.28^{+0.37 }_{-0.20}$ & $0.9^{+1.4}_{-0.0}$ & $-2.58^{+0.44}_{-0.14}$ & $0.091^{+0.097}_{-0.069}$ & $0.34^{+0.00}_{-0.09}$ \\ 
\enddata
\tablenotetext{a}{The source has two possible solutions, and the main one is given in the first entry.}
\tablecomments{Results of the MCMC experiments are obtained with no prior knowledge of BH masses. 
A uniform prior distribution is adopted for the model parameters, including the BH masses. The uncertainties 
of the model parameters are determined by the 90\% confidence level obtained with the MCMC experiments.}
\end{deluxetable*}

\begin{figure}
	\begin{center}
		\includegraphics[width=0.9\columnwidth]{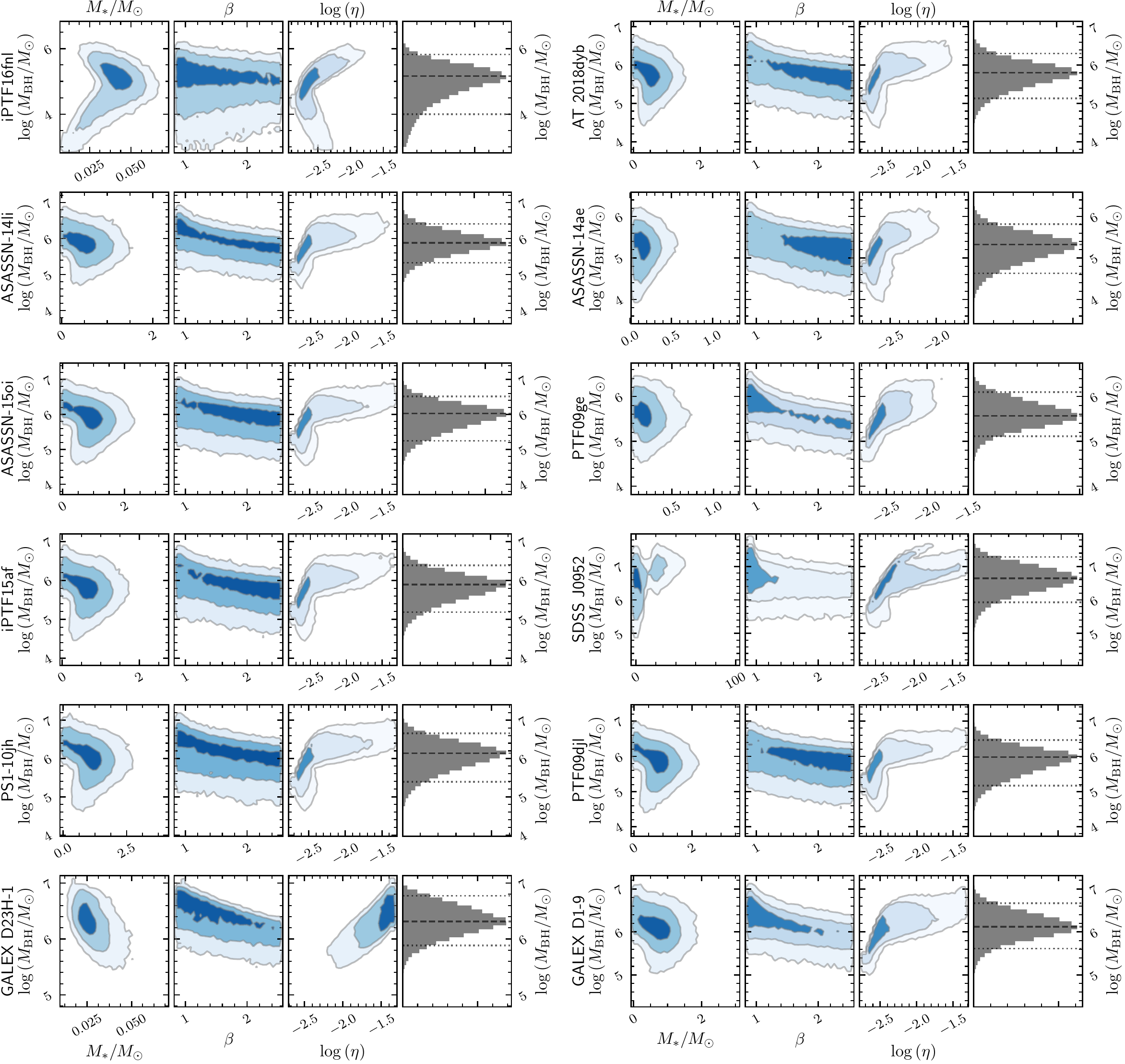}
		\caption{Posterior distributions of the model parameters and the radiation efficiency of the MCMC
experiments. The results are for the TDE sources with observations of stellar velocity dispersion. A uniform prior 
distribution is adopted for all the model parameters ($M_{\rm BH}$, $M_*$ and $\beta$). Contour plots are 
for 1, 2, and $3\,\sigma$. The dashed and dotted lines are the same as in Figure~\ref{fig:contour_mbh}. The main solution is shown when two solutions exist for a TDE.
	\label{fig:mcmc-contour}
		}
	\end{center}
\end{figure}
\begin{figure}
	\begin{center}
		\includegraphics[width=0.9\columnwidth]{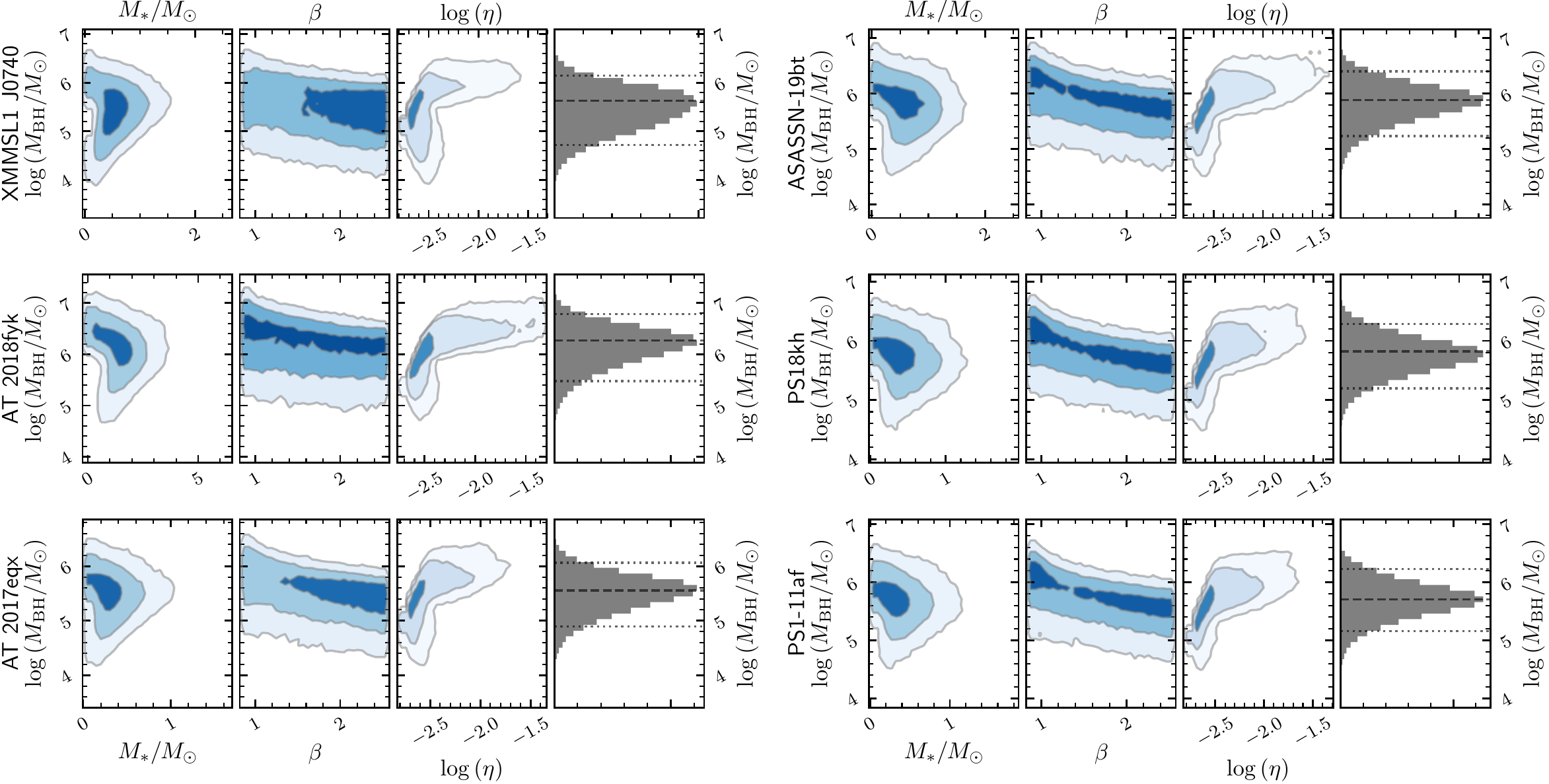}
		\caption{Same as in Figure~\ref{fig:mcmc-contour}, but for the TDE sources lacking
stellar velocity dispersion measurements.
        \label{fig:mcmc-contour2}
		}
	\end{center}
\end{figure}

\begin{figure}
\begin{center}
\includegraphics[width=0.9\columnwidth]{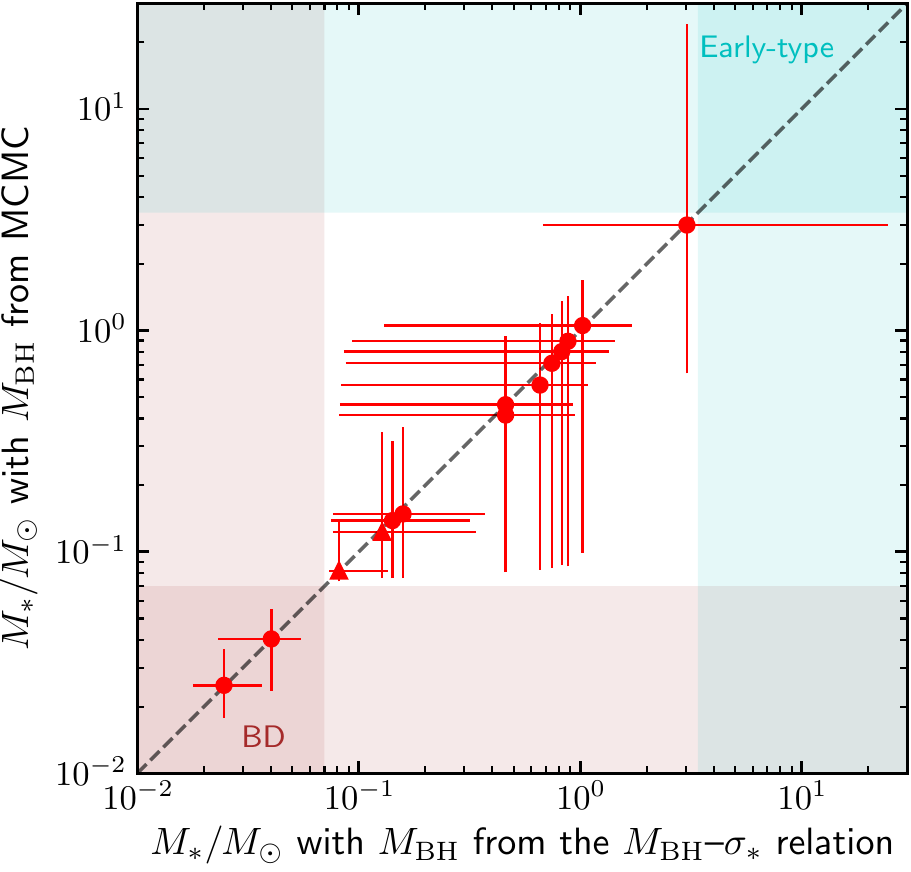}
\caption{Stellar masses of the sample sources with the BH masses estimated from the MCMC experiments in  this paper with those derived from the $M_{\rm BH}$--$\sigma_*$ relation of \citet{vandenBosch2016}.  The filled circles are for the main solutions of TDEs, and the filled triangles are for the secondary solutions.  The uncertainties are determined at the 90\% confidence level in the MCMC experiments. The stellar masses obtained with the two MCMC experiments are consistent with each other.
\label{fig:mstar-compare}
}
\end{center}
\end{figure}

\begin{figure}
\begin{center}
\includegraphics[width=0.9\textwidth]{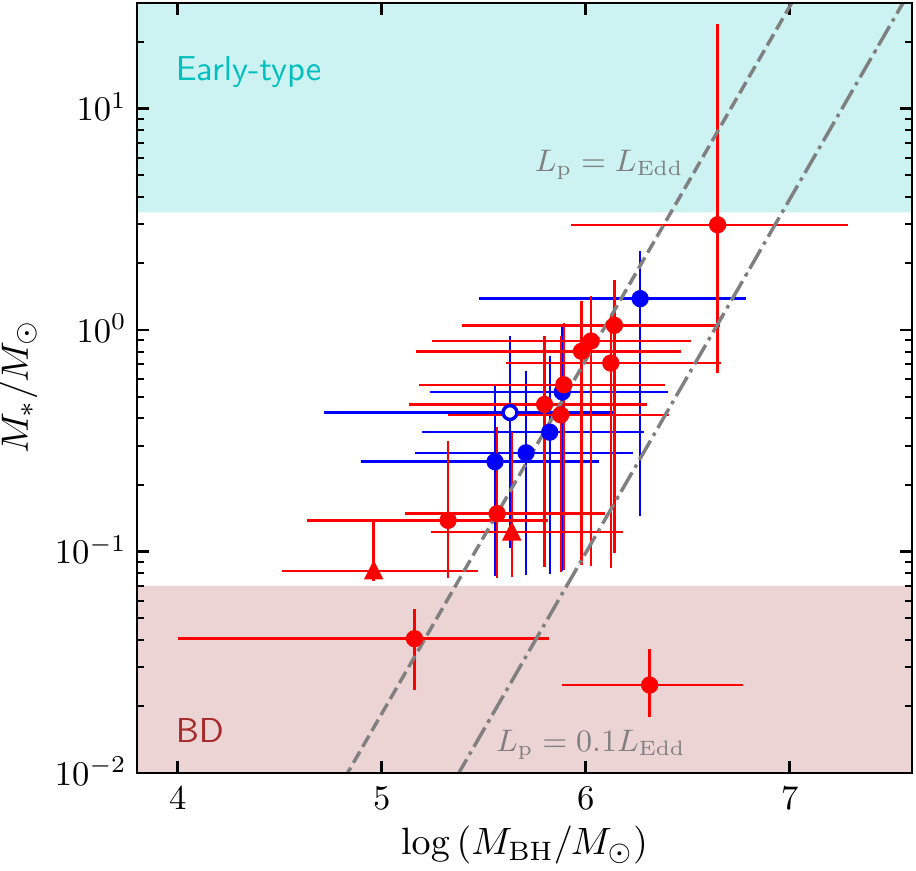} 	
\caption{Stellar mass vs. BH mass for all the sample sources derived from our MCMC experiments.
The red and blue symbols show the results obtained with and without prior knowledge of the BH mass, respectively. The open circle is the X-ray TDE XMMSL1 J0740, and the 
filled triangles are for the secondary solutions of iPTF16fnl and GALEXD23H-1. The dashed and dash--dotted 
lines are the observational limits imposed by Equation~(\ref{eq:seleffect_Edd}) for
$L_{\rm p} = L_{\rm Edd}$ and $0.1\, L_{\rm Edd}$, respectively (see Section~\ref{sec:starmass} for details).
\label{fig:mstar-mbh}
}
\end{center}
\end{figure}

We solve the equations for all the TDEs in Table~\ref{tab:obs} and give the posterior distributions of the
model parameters ($\log{M}_{\rm BH}$, $M_*$, and $\beta$) in 
Figure~\ref{fig:mcmc-contour} and Figure~\ref{fig:mcmc-contour2} for the TDEs with and without the 
observations of the stellar velocity dispersion, respectively.  In Figure~\ref{fig:mcmc-contour} 
and Figure~\ref{fig:mcmc-contour2}, we also give the posterior distributions of the associated radiation 
efficiency $\log\eta$ of the MCMC experiments. When a TDE has two possible solutions with comparable
posterior probabilities, we give the posterior distribution of the primary solution in 
Figure~\ref{fig:mcmc-contour}. Table~\ref{tab:mcmc} shows the resulting masses of the BHs and stars,, 
the radiation efficiency, and the associated uncertainties at the 90\% confidence level. The posterior distributions of the model parameters ($M_{\rm BH}$, 
$M_*$, $\beta$, and $\eta$) shown in Figure~\ref{fig:mcmc-contour} are 
similar to those in Figure~\ref{fig:contour_mbh}, implying that $L_{\rm p}$ and $\Delta{E}$ together can 
determine the mass of the disrupted star, the penetration factor, and the radiation efficiency as well as those when providing the BH masses. In Figure~\ref{fig:mstar-compare} we compare the masses of the disrupted stars derived with and without the knowledge of the BH masses. In the former case, the BH masses are given by the $M_{\rm BH}$--$\sigma_*$ relation.  
It shows that the stellar masses derived in the above two cases are consistent with each other. 
This result suggests that the stellar mass and the amount of the accreted matter can be well constrained by $L_{\rm p}$ and $\Delta{E}$. 
Figure~\ref{fig:mstar-mbh} gives the stellar masses obtained with $L_{\rm 
p}$ and $\Delta{E}$. It shows that the distributions of the stellar types 
of the TDEs with and without the observations of the stellar velocity dispersion are consistent. The stars of the TDE sample sources except iPTF16fnl and GALEX D23H-1 are A- or later-type main-sequence stars, which is consistent with the conclusions obtained by the TDEs with the observed stellar velocity dispersions. 
Figure~\ref{fig:mstar-mbh} shows that the X-ray TDE XMMSL1 J0740 has the highest stellar mass for 
a given BH mass. However, the difference is not significant, and TDE XMMSL1 J0740 is the only 
sample source discovered in the X-ray wave band. Many more X-ray-discovered TDEs
are needed. The correlation between the stellar and BH masses may be due to observational selection effects, as suggested 
by Equation~(\ref{eq:seleffect_Edd}).


\begin{figure} 
	\centering 
	\subfigure{
		\begin{minipage}[b]{0.46\textwidth} 
			\includegraphics[width=1\textwidth]{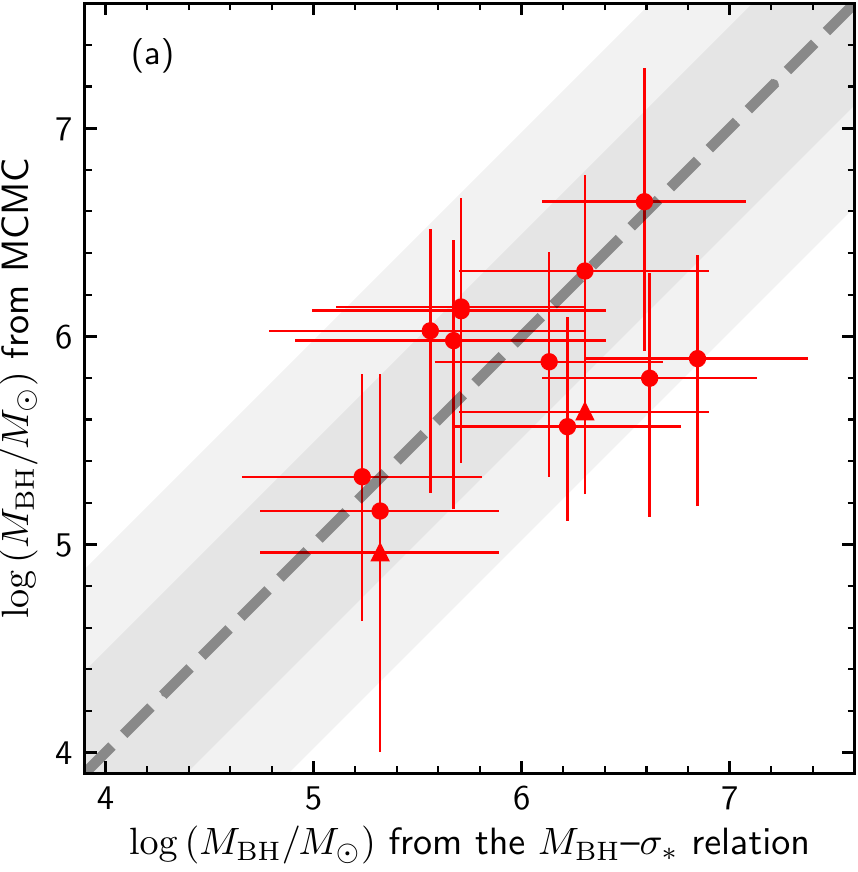} 
			\label{fig:mbh-mbhsigma}
		\end{minipage} 
	}
	\subfigure{
		\begin{minipage}[b]{0.46\textwidth} 
			\includegraphics[width=1\textwidth]{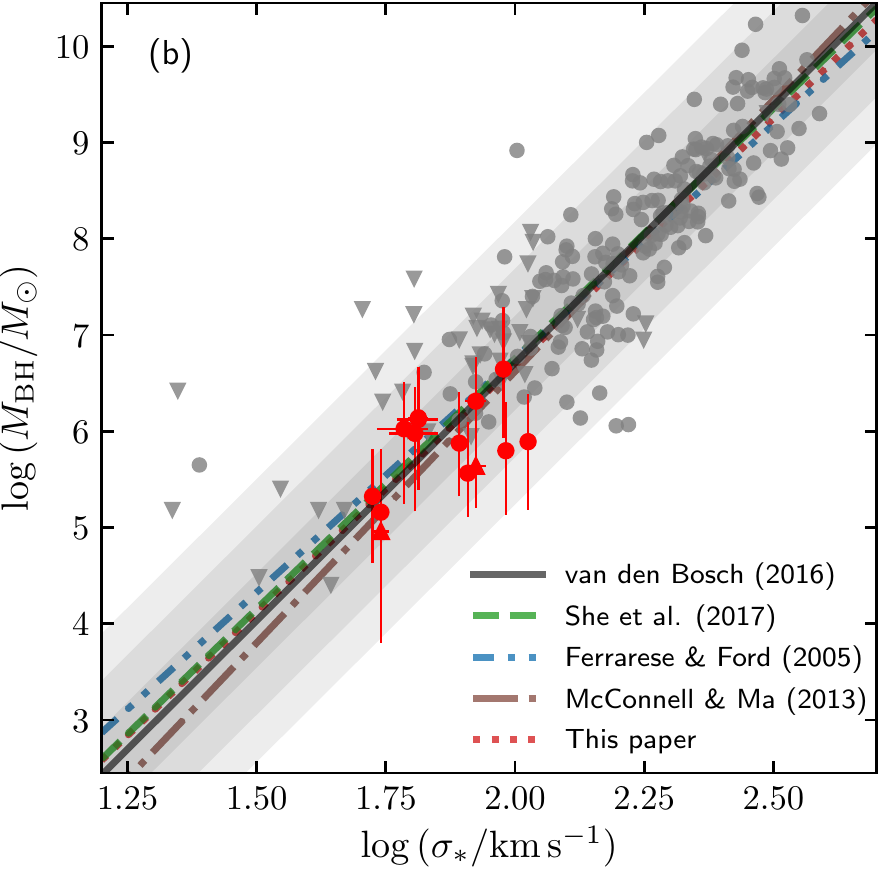} 	
			\label{fig:mbh-sigma}		
		\end{minipage}	
	} 
	\caption{(a) Comparison between the BH mass estimated in this paper with that obtained from the $M_{\rm BH}$--$\sigma_*$ relation. The dashed line shows the one-to-one relation, and the dark and light gray regions denote one and two times the intrinsic scatter of the $M_{\rm BH}$--$\sigma_*$ relation \citep{vandenBosch2016}.  (b) Correlation between BH mass and stellar velocity dispersion. The BH masses obtained in this paper (red points) are overplotted on the $M_{\rm BH}$--$\sigma_*$ relation of \citet{vandenBosch2016}, which is derived from the data (gray points) in his Table~2. Dark to light gray regions denote one, two, and three times the intrinsic scatter. The lines are the $M_{\rm BH}$--$\sigma_*$ relations for all types of galaxies compiled from the literature and the interpolated relation derived in this paper. The BH masses obtained in this paper closely follow the $M_{\rm BH}$--$\sigma_*$ relation used in the literature.  \label{fig:mbh-compare}
	}
\end{figure}

In Figure~\ref{fig:mbh-mbhsigma} we compare the BH masses in Table~\ref{tab:mcmc} obtained with $L_{\rm p}$ and $\Delta{E}$ with those in Table~\ref{tab:obs} calculated with the $M_{\rm BH}$--$\sigma_*$ relation. The 
one-to-one line and the intrinsic scatter of the $M_{\rm BH}$--$\sigma_*$ relation are given to show the expected 
correlation and intrinsic scatter of the BH masses obtained from the two different methods. The uncertainty of the 
BH mass calculated with the $M_{\rm BH}$--$\sigma_*$ relation includes both the observational uncertainty of the 
velocity dispersion and the intrinsic scatter of the $M_{\rm BH}$--$\sigma_*$ relation. Figure~\ref{fig:mbh-mbhsigma} 
shows that the BH masses obtained with $L_{\rm p}$ and $\Delta{E}$ for all the TDEs except iPTF15af are consistent 
within one sigma with the BH masses obtained from the $M_{\rm BH}$--$\sigma_*$ relation. The BH mass of iPTF15af 
computed from $L_{\rm p}$ and $\Delta{E}$ is lower by $0.97$ dex or $1.8$ times the standard deviation 
($0.53$ dex) than the mass obtained with the $M_{\rm BH}$--$\sigma_*$ relation. The UV spectra of TDE iPTF15af have 
broad absorption lines associated with high-ionization states of \ion{N}{5}, \ion{C}{4}, \ion{Si}{4}, and possibly 
\ion{P}{5}. These features require an absorber with column densities $N_{\rm H} > 10^{23}\, {\rm cm^{-2}}$ 
\citep{Blagorodnova2019}. Such an optically thick gas could significantly absorb the soft X-rays, if present. However, 
the observations of soft X-rays in the optically discovered TDEs suggested that the radiation in soft X-rays is much lower than or at most comparable to that in the optical and UV wave bands. Therefore, the low value of the BH 
mass of iPTF15af obtained with $L_{\rm p}$ and $\Delta{E}$ may not mainly be due to the absorption of soft X-rays, but to the intrinsic scatter of the $M_{\rm BH}$--$\sigma_*$ relation. In Figure~\ref{fig:mbh-sigma} we overplot the BH 
masses of the TDEs obtained with $L_{\rm p}$ and $\Delta{E}$ on the $M_{\rm BH}$--$\sigma_*$ relation obtained by
\citet{vandenBosch2016}. The data are adopted from Table~2 of \citet{vandenBosch2016}, in which the BH masses
are derived from stellar dynamics, gas dynamics, megamasers, and reverberation mapping. 
\citet{Kormendy2013} carefully refined all the present observational data, but only provided an 
updated $M_{\rm BH}$--$\sigma_*$ relation for the galaxies with elliptical and classical bulges.  The $M_{\rm BH}$--$\sigma_*$ relation for all galaxies with those tabulated data has been given only recently \citep{She2017}. The two
formulations of the $M_{\rm BH}$--$\sigma_*$ relation for all galaxies obtained both by \citet{She2017} and by \citet{vandenBosch2016} are shown in Figure~\ref{fig:mbh-sigma} and are nearly identical to each other, 
justifying  the results calculated based on the $M_{\rm BH}$--$\sigma_*$ relation obtained by \citet{vandenBosch2016}. 
For comparison, Figure~\ref{fig:mbh-sigma} also shows several popular $M_{\rm BH}$--$\sigma_*$ relations, which 
are obtained for all types of galaxies \citep{Ferrarese2005,McConnell2013} and were recently used to estimate 
the BH masses of TDEs \citep{Stone2016,Blagorodnova2017,Wevers2017,Wevers2019,Leloudas2019}. 
Figure~\ref{fig:mbh-sigma} shows that the $M_{\rm BH}$--$\sigma_*$ relations for all galaxies are well consistent
with one another. The BH masses obtained in this work are located in the core regions of the correlation, with a scatter 
comparable to the intrinsic scatter of the $M_{\rm BH}$--$\sigma_*$ relation. In Figure~\ref{fig:mbh-sigma}, the 
interpolated $M_{\rm BH}$--$\sigma_*$ relation from Equation~(\ref{eq:mbhsigma}) is also shown and remarkably 
consistent with those for all types of galaxies in the literature as well as with the BH masses obtained in this work.

In Section~\ref{sec:obs} we showed observationally and theoretically that the bolometric luminosity and the 
total radiation energy could properly include the EUV radiation by integrating over a single blackbody from the optical 
and UV radiation and adding the observations of soft X-ray wave bands. The consistencies of the BH masses obtained
in this paper and with the $M_{\rm BH}$--$\sigma_*$ relation also suggest that the conclusions are reasonable.
However, the spectral energy distributions of a few TDEs occasionally deviate from the single-temperature blackbody spectrum, and the contribution of EUV radiation in the bolometric luminosity and
the total radiation energy cannot be well constrained until direct observations of EUV radiation are available. Here we briefly
discuss the effects of the EUV radiation on the results by arbitrarily increasing by 0.5~dex the peak bolometric
luminosity $L_{\rm p}$ and the total radiation energy $\Delta{E}$ of the well-known PS1-10jh in Table~\ref{tab:obs}. Such an operation is equivalent to the assumption that the EUV radiation is about 5 times the observed optical/UV 
radiation and the color index does not significantly change with time. We note that the hypothetical
bolometric peak luminosity $L_{\rm p} = 10^{44.84} \, {\rm erg\,s^{-1}}$ is about 11~times the Eddington luminosity 
for the BH mass $\log(M_{\rm BH}/M_\odot)=5.71$ given by the $M_{\rm BH}$--$\sigma_*$ relation in Table~\ref{tab:obs}, 
and this luminosity should lead to a top-capped light curve due to the Eddington limit, but this theoretical prediction is inconsistent with the observation of PS1-10jh \citep{Gezari2012}. Here we 
neglect the inconsistence and investigate the effects of the possibly missed EUV radiation on the results. With 
the arbitrarily assumed bolometric peak luminosity $L_{\rm p}$ and total radiation energy $\Delta{E}$, we solve Equations~(\ref{eq:plum}) and (\ref{eq:toteng}) with the MCMC method. The results suggest a BH mass of 
$\log(M_{\rm BH}/M_\odot)=6.39_{-0.82}^{+0.74}$, a stellar mass of $M_*/M_\odot=2.87_{-2.44}^{+18.75}$, and 
a radiative efficiency of $\log(\eta) = -2.46^{+0.50}_{-0.16}$. These values indicate that a significant increase in
EUV radiation from about one to about five times the observed value in optical/UV would increase the radiation
efficiency only by 0.11~dex, the BH mass by $0.25$~dex, and the mass of the star from $1.05\,M_\odot$ to $2.87\,M_\odot$.
A star of mass $M_*/M_\odot=2.87$ is an A-type main-sequence star and is roughly consistent with the 
constraint from the star-formation history of the host galaxy of PS1-10jh \citep[Figure~1 of][]{French2017}. In addition, an increase in peak bolometric luminosity and total 
radiation energy of PS1-10jh by 0.5~dex would lead to a moderate increase in measured BH mass 
by 0.25~dex. The BH mass $\log(M_{\rm BH}/M_\odot)=6.39$ is consistent within $2\sigma$ with the BH mass $
\log(M_{\rm BH}/M_\odot)= 5.71_{-0.60}^{+0.59}$ obtained with the $M_{\rm BH}$--$\sigma_*$ relation. These results imply that the EUV radiation, if significant, would not change our conclusions.

\subsection{BH masses from the $M_{\rm BH}$--$M_{\rm bulge}$ relation}
\label{sec:mbh-compare}

\begin{figure}
	\begin{center}
		\includegraphics[width=0.9\columnwidth]{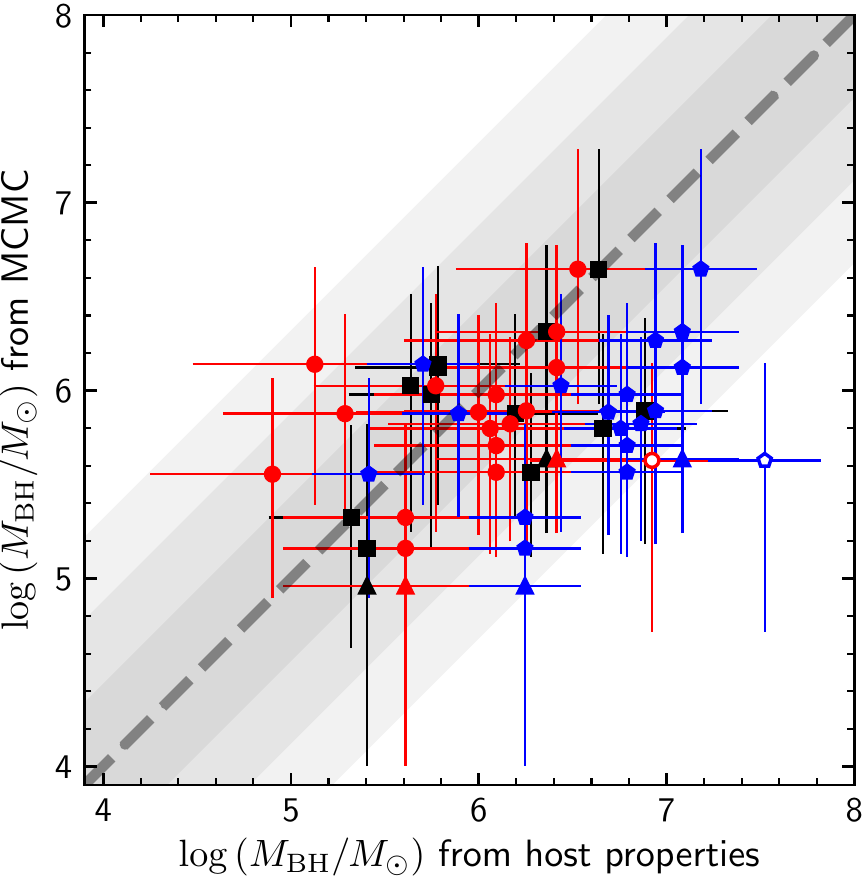}
		\caption{Comparison between the BH masses obtained in this paper 
with those estimated from the total and bulge masses of the host galaxies. The filled pentagons are obtained with the 
$M_{\rm BH}$--$M_{\rm bulge}$ relation, while the filled circles come from the $M_{\rm BH}$--$M_{\rm tot}$ relation.  The 
open pentagon and open circle are for the X-ray TDE XMMSL1 J0740.  The BH masses calculated with the $M_{\rm 
BH}$--$\sigma_*$ relation \citep{She2017} are overplotted for comparison as filled squares. The filled triangles are for 
the secondary solutions of iPTF16fnl and GALEXD23H-1. The dashed line is the one-to-one relation, and the dark to 
light gray regions give one, two, and three times the intrinsic scatter of the $M_{\rm BH}$--$\sigma_*$ relation of 
\citet{She2017}. } 
			\label{fig:mbh-compare2}		
	\end{center}
\end{figure}

In Section~\ref{sec:method} we computed the BH masses with $L_{\rm p}$ and $\Delta{E}$ and showed
that they are consistent with the BH masses calculated with the $M_{\rm BH}$--$\sigma_*$ relation. 
In addition to the $M_{\rm BH}$--$\sigma_*$ relation, the BH masses can also be calculated with the bulge masses
$M_{\rm bulge}$ of the host galaxies \citep{Magorrian1998}. The 
$M_{\rm BH}$--$M_{\rm bulge}$ relation for classical bulges and ellipticals has the same intrinsic scatter as
the $M_{\rm BH}$--$\sigma_*$ relation \citep{Haering2004,Kormendy2013}. However, it has been noted in the literature 
that the BH masses of TDEs derived with the $M_{\rm BH}$--$M_{\rm bulge}$ relations are systematically higher 
than those obtained with the $M_{\rm BH}$--$\sigma_*$ relation \citep{Gezari2017,Wevers2017,Mockler2019}, 
and the BH masses of AGNs are also an order of magnitude lower than those calculated with the $M_{\rm BH}$--$M_{\rm bulge}$ relation. There are several possible explanations for the discrepancy: (1) TDEs are expected 
to occur in dwarf galaxies and it is difficult to spatially resolve the bulges of the host galaxies, and (2) the host galaxies of most TDEs are E+A galaxies or post-starburst 
galaxies \citep{Arcavi2014,French2016}, which are in transition between late-type spirals and passive early-type
galaxies and have overdense centers with respect to the galaxies from which the $M_{\rm BH}$--$M_{\rm bulge}$ relation is derived \citep{French2017}. It has recently been suggested that the BH masses of TDEs obtained with the 
$M_{\rm BH}$--$M_{\rm bulge}$ and $M_{\rm BH}$--$\sigma_*$ relations may be roughly consistent with each other when the $M_{\rm BH}$--$M_{\rm bulge}$ relation for all types of host galaxies are used and the $B/T$ ratio is estimated from the total stellar mass and averaged over all types of galaxies \citep{Wevers2019}. Here we follow this approach to estimate the BH masses. We estimate the $B/T$ ratio 
for our sample sources using the empirical relation between the total stellar mass of the host galaxy and the 
averaged $B/T$ obtained for all types of galaxies \citep{Stone2018}. The resulting $B/T$ ratios are given in Table~\ref{tab:obs} and have very large uncertainties. Following  \citet{Wevers2019}, we 
estimate the BH masses using the $M_{\rm BH}$--$M_{\rm bulge}$ relation for all types of galaxies 
\citep{Haering2004}. In Figure~\ref{fig:mbh-compare2} we overplot the BH masses obtained with the $M_{\rm BH}$--$M_{\rm bulge}$ relation on the $M_{\rm BH}$--$M_{\rm BH}$ plot. Because it is difficult to estimate the
uncertainties of the total stellar masses and the $B/T$ ratios, we only show the uncertainty of the $M_{\rm BH}$--$M_{\rm bulge}$ relation in Figure~\ref{fig:mbh-compare2}. The real scatter in the BH mass should be larger than what is shown here. Figure~\ref{fig:mbh-compare2} 
shows that the BH masses with the  $M_{\rm BH}$--$M_{\rm bulge}$ relation are largely consistent with the BH masses 
obtained from $L_{\rm p}$ and $\Delta{E}$ in this paper and from the $M_{\rm BH}$--$\sigma_*$ relation, 
although a small systematic difference is possible.

It is well known that the BHs in quiescent galaxies do not correlate with the galaxy disks \citep[][and references 
therein]{Kormendy2013}. It has recently been shown that the BH masses of local AGNs may correlate with the total 
stellar mass of the host galaxies $M_{\rm tot}$ \citep{Reines2015}. 
The relation between BH masses and total stellar masses of the host galaxies of AGNs is used to estimate the BH 
masses of TDEs in the literature, especially when the stellar velocity dispersion of the host is not available 
\citep[e.g.,][]{Komossa2004,Gezari2017,Lin2017a,Lin2017}. Using the updated $M_{\rm BH}$--$M_{\rm
tot}$ relation \citep{Greene2020}, we calculate the BH masses of the TDE sources with the total stellar masses 
shown in Table~\ref{tab:obs} and give the results in Table~\ref{tab:obs}. We overplot the results on the $M_{\rm 
BH}$--$M_{\rm BH}$ plot in Figure~\ref{fig:mbh-compare2}. It shows that the BH masses with the  $M_{\rm 
BH}$--$M_{\rm tot}$ relation are well consistent with both the BH masses obtained in this paper and with the 
$M_{\rm BH}$--$\sigma_*$ relation.

\subsection{BH masses obtained by fitting the light curves of TDEs}
\label{sec:mbh-lcurve}

In this paper we propose a method of measuring the masses of BHs and stars of TDEs by jointly fitting the peak bolometric luminosity and the total radiation energy. In the calculations, we use a mass-to-radiation conversion efficiency that is computed from the elliptical disk model suggested by \citet{Liu2017}. \citet{Mockler2019} recently proposed measuring
the BH masses by fitting the observed light curves of TDEs using the outputs from the numerical simulations of the fallback rate for stellar debris \citep{Guillochon2013}. In their calculations, the conversion efficiency of matter into radiation is assumed to be
a free parameter of agnostic physics origin \citep{Mockler2019}. 

\begin{figure}
\begin{center}
\includegraphics[width=0.9\columnwidth]{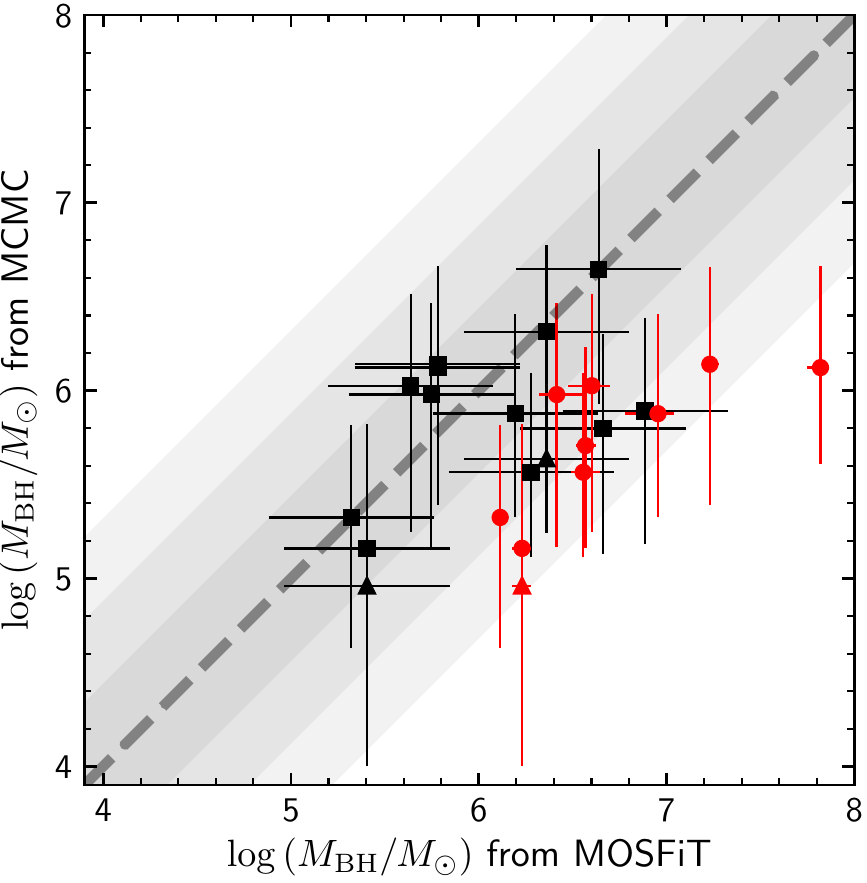}
\caption{Comparison of the BH masses obtained in this paper with those derived by fitting the 
multiwavelength light curves of TDEs \citep{Mockler2019}. BH masses calculated with the 
$M_{\rm BH}$--$\sigma_*$ relation \citep{She2017} are overplotted for comparison as filled squares. 
The dashed line is the one-to-one relation, and the dark to light gray regions give one, two, and three times the
intrinsic scatter of the $M_{\rm BH}$--$\sigma_*$ relation of \citet{She2017}.
\label{fig:mbh-TDEs}
}
\end{center}
\end{figure}

Figure~\ref{fig:mbh-TDEs} plots the BH masses obtained in this work and in \citet{Mockler2019}. 
We do not compare the stellar masses because the method of \citet{Mockler2019} cannot constrain 
the stellar mass due to the strong degeneracy of the stellar 
mass and the orbital penetration factor. We also plot in Figure~\ref{fig:mbh-TDEs} the BH 
masses obtained with the $M_{\rm BH}$--$\sigma_*$ relation for comparison. The plot shows 
that the BH masses obtained by \citet{Mockler2019} are roughly consistent with but systematically higher than those obtained with 
$L_{\rm p}$ and $\Delta{E}$ and the $M_{\rm BH}$--$\sigma_*$ relation. It has already been noted in the literature that the method of fitting the 
light curves systematically produces higher BH masses than the $M_{\rm BH}$--$\sigma_*$ relation 
\citep{Mockler2019}. Of all the sample sources, GALEX D1-9 is the most controversial because the BH masses derived from the three methods are very different. 
The light curve of GALEX D1-9 suggests a BH mass about $6.6\times 10^7 \,M_\odot$ \citep{Mockler2019}, which is 
about two orders of magnitude larger than that ($10^{5.71} \,M_\odot$) obtained with the $M_{\rm BH}$--$\sigma_*$ relation and about fifty times larger than the result $10^{6.12} \,M_\odot$ of this
paper.

\section{Discussion} 
\label{sec:dis}

Liu and collaborators recently suggested that the accretion disks of TDEs are extended and highly eccentric 
with nearly uniform eccentricity. The nearly constant orbital eccentricity of the disk fluid elements 
during the accretion onto the BH could explain the complex profiles of the observed emission lines
\citep{Liu2017,Cao2018}. Here we calculated the radiation efficiency of the elliptical accretion disk model.
Our results show that the radiation efficiency of the highly eccentric accretion disk depends on the masses of the BHs and stars as well as on the orbital penetration factors of the stars. The values could significantly vary with TDEs. The radiation efficiency of 
the elliptical accretion disk model could be as low as $10^{-3}$, or about two orders of magnitude lower than the 
typical radiation efficiency $\eta = 0.1$ adopted for TDEs in the literature. Based on the elliptical accretion disk 
model, we calculate the expected peak luminosity and total radiation energy after peak, which can be well
determined by observations of TDEs. 

We compile from the literature the observational data of the peak bolometric luminosities and total radiation energies 
after peak for a sample of 18 non-jetted TDEs in quiescent galaxies. Twelve of these TDEs 
have available stellar velocity dispersions from the observations of their host galaxies, so that the BH masses can be calculated from the $M_{\rm BH}$--$\sigma_*$ relation. We show that the peak bolometric luminosities and the total radiation energies computed from the elliptical disk model are consistent with the observational data. The low peak luminosity
and apparently low accreted mass could be explained by the unusually low radiation efficiency of elliptical accretion 
disks without requiring alternative explanations for the transient sources \citep[e.g.,][]{Saxton2018} or missing the majority of the
released energy in the EUV \citep[e.g.,][]{Lu2018}. 

Given $L_{\rm p}$ and $\Delta{E}$, we can calculate the radiation efficiency and determine the mass of the disrupted star through Equations~(\ref{eq:plum}) and (\ref{eq:toteng}) using the MCMC experiments, regardless of a prior knowledge of the BH mass. 
Our sample sources except for GALEX D23H-1 have a typical radiation efficiency $\eta\simeq 2.7\times 10^{-3}$, which is about 37 times 
lower than the typical radiation efficiency $\eta=0.1$ adopted for TDEs in the literature. Our results are consistent with those from earlier work. The
radiation efficiency of PTF09djl and ASASSN-14li are $ \log{\eta}= 
-2.61^{+0.58}_{-0.10}$ and $\log{\eta} = -2.57^{+0.54}_{-0.14}$, respectively. These values are consistent with the results $\log{\eta}\simeq -2.38$ for PTF09djl and $\log{\eta}\simeq -2.43$ for ASASSN-14li, obtained by modeling the profiles of the broad optical emission lines
\citep{Liu2017,Cao2018}.  The radiation
efficiencies of a sample of TDEs have recently been obtained by fitting the light curves with the fallback rate of stellar 
debris \citep[MOSFiT,][]{Mockler2019}. However, the uncertainties of the MOSFiT results are very large because the 
radiation efficiency from the MOSFiT method is strongly degenerate with stellar masses ranging from $0.01 
\,M_\odot$ to $100 \,M_\odot$ and cannot be determined uniquely. As an example, Table~5 of \citet{Mockler2019} 
gave the radiation efficiencies of TDE PS1-10jh obtained with different stellar masses. They showed that the values
change from $\eta=0.9\times 10^{-1}$ for $M_* = 0.1\,M_\odot$,  through $\eta=3.8\times 10^{-3}$ for $M_* = 1.0 
\,M_\odot$, to $\eta= 4\times 10^{-4}$ for $M_* = 10 \,M_\odot$. No strong prior can be given to the stellar mass, and 
the uncertainty is as large as 3 dex. This uncertainty is much larger than the quoted uncertainty of the fiducial value $\eta=0.09^{+0.03}_{-0.02}$ 
\citep{Mockler2019}. In this paper, the stellar mass and radiative efficiency of the TDE PS1-10jh can be determined separately. The results are shown in Table~\ref{tab:mcmc_mbh}, which are $M_* = 1.02^{+0.69}_{-0.89} \,M_\odot$ and $\log(\eta) =
-2.55^{+0.68}_{-0.13}$. Our result is significantly smaller than the fiducial one from the MOSFiT method. Interestingly, 
our result is consistent with the test result of $M_* = 1.0 \,M_\odot$ and $\eta=3.8\times 10^{-3}$ (or 
$\log{\eta}=-2.42$) of the MOSFiT method. Taking into account that the systematic uncertainty of the result of the MOSFiT method is large, we conclude that the radiation efficiencies obtained in this paper are consistent with those from the MOSFiT method.  
	
To produce the same total radiation energy, a low radiation efficiency requires a large amount of accreted  matter onto the BH. We calculated the amount of accreted stellar matter after peak and showed that
it is in the range of about $10^{-2} \,M_\odot$ to
$0.97 \,M_\odot$, about 34\% of the mass of the star. The fraction of 34\% is the expectation of hydrodynamic 
simulations of tidal disruption of low-mass stars with a polytropic index $\gamma=5/3$ and orbital penetration factor 
$\beta \sim 1$. Our model is unable to constrain the accreted matter before the peak, and we cannot estimate 
the total accreted stellar matter of TDEs. The results imply that most of the orbital energy of the stellar debris 
is advected onto the BH instead of being converted into radiation in the EUV, which last is often assumed in the literature
\citep[e.g.,][]{Lu2018}. In addition, because of the low peak bolometric luminosity and total radiation energy as well as the apparently low 
 accreted matter of the optical/UV nuclear transients, we do not require an alternative explanation, as suggested in 
 \citet{Saxton2018}.

 We also find that the disrupted stars of our TDE sample except SDSS J0952+2143 are in the mass range of $M_* \simeq 
 2.5\times 10^{-2} \,M_\odot$ and $1.4 \,M_\odot$. The spectral types range from  brown dwarfs to late A-type main 
sequence. The absence of B- and O-type stars in our sample is consistent with the observational fact that 
the host galaxies of many TDEs are E+A galaxies with a burst of star formation about a few billion years ago 
\citep{Arcavi2014,French2016}. The stellar mass of the TDE in SDSS J0952+2143 is about $3.0 \,M_\odot$ and 
consistent with the fact that the host galaxy of SDSS J0952+2143 is a star-forming galaxy \citep{Komossa2008b} rich in 
young stars \citep{Palaversa2016}.

With the peak bolometric luminosity and the total radiation energy after peak, we can constrain the masses of the BHs of TDEs. The BH masses obtained 
in this paper are consistent with those obtained with the classical $M_{\rm BH}$--$\sigma_*$, $M_{\rm BH}$--$M_{\rm 
bulge}$, and $M_{\rm BH}$--$M_{\rm tot}$ relations. It was noted in the literature that the BH masses of TDEs 
given with the $M_{\rm BH}$--$L_{\rm bulge}$ or $M_{\rm BH}$--$M_{\rm bulge}$
relation are much higher than those calculated with the $M_{\rm BH}$--$\sigma_*$ relation. Our results suggest
that the discrepancy is most probably due to the difficulty of accurately measuring the mass of the host galaxy bulge.

The BH masses of many of our sample TDE sources have recently been calculated with the MOSFiT method, which is based 
on analyzing the multiwave-band light curves \citep{Mockler2019}. The BH masses of the TDEs obtained in this paper 
and with MOSFiT are largely consistent with each other, with some exceptions. For GALEX D1-9 and PS1-10jh, the 
MOSFiT method gives $M_{\rm BH}= 6.6 \times 10^7 \,M_\odot$ and $M_{\rm BH}= 1.7 \times 10^7 
\,M_\odot$, respectively, which is about 50 times and 12 times higher, respectively, than our results. It has also been noted in 
the original work \citep{Mockler2019} and in this paper that the measurements of the BH masses of GALEX D1-9 and 
TDE PS1-10jh with MOSFiT are much higher than the measurement of $10^{5.71}\,M_\odot$ calculated 
with the $M_{\rm BH}$--$\sigma_*$ relation. The BH mass in the MOSFiT method is determined under the assumption that 
the bound debris is promptly circularized, so that the luminosity closely follows the mass fallback rate without significant 
delay \citep{Mockler2019}. A prompt circularization of stellar debris streams and the rapid formation of an accretion 
disk are expected only for tidal disruptions of the stars with an orbital pericenter of  about the gravitational radius of the 
BH \citep{Dai2015,Shiokawa2015,Bonnerot2016,Hayasaki2016a}. The circularization of debris streams is slow for BHs 
of mass $\sim 10^6\,M_\odot$ and a typical penetration factor $\beta \sim 1$ because the general relativistic apsidal 
precession is weak. An inefficient circularization of stellar debris would result in a rise-to-peak timescale longer than 
that given by $\Delta{t}_{\rm p}$. A direct comparison of the observed rise-to-peak timescale and the expected 
$\Delta{t}_{\rm p}$ would require a more massive BH solution of the MOSFiT method.

Twelve of the 18 TDEs in our sample have stellar velocity dispersions measured from the host 
galaxies. When the BH masses obtained in this paper are plotted on the $M_{\rm BH}$--$\sigma_*$ 
diagram with the BH masses measured by the stellar dynamics, gas dynamics, megamasers, and 
reverberation mapping, we find that they share the same $M_{\rm BH}$--$\sigma_*$ 
relation and that the BH masses in our sample distribute in the low-mass region of the 
$M_{\rm BH}$--$\sigma_*$ diagram. The results indicate that the method in this paper can give an independent 
accurate measurement of the BH mass and test the $M_{\rm BH}$--$\sigma_*$ 
relation at low BH mass. We have calculated the BH masses of the 6 sample TDEs without 
measuring the stellar velocity dispersions of the host galaxies. It would be interesting to compare these masses with those derived in the future from the $M_{\rm BH}$--$\sigma_*$ relation. 

The masses of the BHs and the tidally disrupted stars obtained in this paper are determined
mainly by the absolute peak brightness and the total radiation energy integrated over the light curve 
after peak. The results depend very weakly on either the shapes of the light curves or the properties of the stars. The consistency between the BH masses obtained in this paper and those from the $M_{\rm BH}$--$\sigma_*$ relation, as well as the agreement between the accreted stellar masses derived from our model and those from the hydrodynamical simulations, justifies the elliptical accretion 
disk model of large size and invariant extreme eccentricity \citep{Liu2017,Cao2018}. Such a disk usually has a sub-Eddington luminosity for the BHs of mass $\ga 10^6 
\,M_\odot$ and is significantly super-Eddington only for tidal disruption of main-sequence stars by intermediate-mass BHs with mass $M_{\rm BH} \la 10^{5} \,M_\odot$. An accretion disk of sub-Eddington luminosity is 
cool and radiatively efficient, and the luminosity closely follows the fallback rate of the stellar debris. The radiation energy of the elliptical accretion disk model is consistent with the 
observations, and no optically thick envelope is needed. No strong disk wind or outflow is 
expected to form on the surface of the cool sub-Eddington accretion disk. However, a small fraction of the fallback 
matter may become unbound and form outflows due to the shocks when the streams collide at the apocenter 
of the elliptical disk \citep{Jiang2016a}. In this case, absorption lines may be detected in the spectrum.

\section{Conclusions}
\label{sec:con}

We have calculated the radiation efficiency for a sample of TDEs based on the elliptical accretion disk model and 
investigated its implications for the observations of TDEs. We showed that the low peak bolometric luminosity and 
low total radiation energy of TDEs result from the low radiation efficiency of the elliptical accretion disk and that 
the main radiation comes from the disk rather than from the self-crossing shocks at apocenter. When the peak 
bolometric luminosity and the total radiation energy after peak are known, we can derive the masses of the BHs and stars.

Since the method in this paper does not require the knowledge of the properties of the host galaxies,  it can also be 
used to measure the masses of off-center recoiling BHs or a component of supermassive BH pairs in galaxy mergers, 
the masses of intermediate-mass BHs in globular star clusters, or the masses 
of primordial BHs wandering in galactic disks or halos. This alternative method for estimating the BH mass is also important for classical galaxies with central supermassive BHs because (1) in dwarf galaxies, accurate $\sigma_*$ 
measurements require deep exposures with very high spectral resolution, (2) measuring $\sigma_*$  
becomes more difficult for high-z sources, and (3) upcoming and next-generation sky surveys are expected 
to detect thousands of TDEs, making spectroscopic follow-ups of all these sources extremely challenging. Finally, we would like to emphasize again that the BH-galaxy scaling relationship below a BH mass of $\sim 10^6 \,M_\odot$ has not been fully explored in the past. TDEs offer a rare opportunity to probe this unexplored regime, as has been demonstrated in this work.


\section*{Acknowledgements}

We would like to thank Hua Gao, Julian Krolik, Nadejda Blagorodnova, Richard Saxton, and Thomas Wevers for 
helpful discussions. We are grateful to the anonymous referee for very helpful comments. This work is supported by 
the National Natural Science Foundation of China (NSFC No.11473003, NSFC No.11721303) and the Strategic Priority 
Research Program of the Chinese Academy of Sciences (grant No. XDB23010200 and No. XDB23040000). L.C.H was
supported by the National Key R\&D Program of China (2016YFA0400702). X.C. acknowledged the support of  the 
National Natural Science Foundation of China (NSFC No.11991053)

\software{
	Astropy \citep{AstropyCollaboration2013}, emcee \citep{Foreman-Mackey2013},
	lmfit \citep{Newville2014}, Matplotlib \citep{Hunter2007}, NumPy \citep{Harris2020}, 
	Pandas \citep{McKinney2010} and SciPy \citep{Virtanen2020}.}

\vskip 1pc



\end{CJK*}
\end{document}